\documentclass[11pt, draftcls, onecolumn]{IEEEtran}

\usepackage{amsmath,amssymb,mathrsfs,graphicx}
\usepackage{bbold}
\usepackage{psfrag}
\usepackage{nccmath}
\usepackage[ulem=normalem]{changes}
\usepackage{algpseudocode,algorithmicx}
\usepackage{algorithm}

\let\oldbrace\{
\def\{{\oldbrace\kern0.5pt}

\newcommand{\nn}{\nonumber}

\newcommand{\Ac}{\mathcal{A}}

\newcommand{\Nc}{\mathcal{N}}

\newcommand{\Qc}{\mathcal{Q}}
\newcommand{\Rc}{\mathcal{R}}
\newcommand{\Sc}{\mathcal{S}}
\newcommand{\Tc}{\mathcal{T}}

\newcommand{\Xc}{\mathcal{X}}
\newcommand{\Yc}{\mathcal{Y}}

\newcommand{\rv}{\mathbf{r}}

\newcommand{\aep}{{\mathcal{T}_{\epsilon}^{(n)}}}

\newcommand{\aepk}{{\mathcal{T}_{\epsilon}^{(k)}}}
\newcommand{\aepvark}{{\mathcal{T}_{\epsilon'}^{(k)}}}

\newcommand{\Rh}{{\hat{R}}}

\newcommand{\Ut}{{\tilde{U}}}
\newcommand{\Vt}{{\tilde{V}}}

\newcommand{\ut}{{\tilde{u}}}
\newcommand{\vt}{{\tilde{v}}}

\def\d{\delta}
\def\e{\epsilon}

\DeclareMathOperator\E{\sf E}
\let\P\relax
\DeclareMathOperator\P{\sf P}

\newcommand{\Bern}{\mathrm{Bern}}

\newcommand{\U}{\mathrm{Unif}}

\DeclareMathOperator*{\argmax}{\arg\max}

\newtheorem{theorem}{Theorem}

\newtheorem{remark}{Remark}

\newtheorem{proposition}{Proposition}
\newtheorem{corollary}{Corollary}

\newcommand{\cl}{{\mathsf{cl}}}
\newcommand{\Rcache}{R_{\sf c}}
\newcommand{\tdash}{{\text{-}}}

\begin{document}
\title{Information Theoretic Caching:\\ The Multi-User Case}

\author{
\IEEEauthorblockN{Sung Hoon Lim, Chien-Yi Wang, and Michael Gastpar} \\
\thanks{S.~H.~Lim and M. Gastpar is with the School of Computer and Communication Sciences, EPFL,
Lausanne, 1015, Switzerland (e-mail:
\{sung.lim, michael.gastpar\}@epfl.ch).}%
\thanks{C.-Y.~Wang is with the Department of Communications and Electronics, Telecom ParisTech, Paris, France (e-mail: chien-yi.wang@telecom-paristech.fr).}%
}


\maketitle

\allowdisplaybreaks
\IEEEpeerreviewmaketitle

\begin{abstract}
In this paper, we consider a cache aided network in which each user is assumed to have individual caches, while upon users' requests, an update message is sent though a common link to all users. First, we formulate a general information theoretic setting that represents the database as a discrete memoryless source, and the users' requests as side information that is available everywhere except at the cache encoder. The decoders' objective is to recover a function of the source and the side information. By viewing cache aided networks in terms of a general distributed source coding problem and through information theoretic arguments, we present inner and outer bounds on the fundamental tradeoff of cache memory size and update rate. Then, we specialize our general inner and outer bounds to a specific model of content delivery networks: File selection networks, in which the database is a collection of independent equal-size files and each user requests one of the files independently. For file selection networks, we provide an outer bound and two inner bounds (for centralized and decentralized caching strategies). For the case when the user request information is uniformly distributed, we characterize the rate vs. cache size tradeoff to within a multiplicative gap of $4$.
By further extending our arguments to the framework of Maddah-Ali and Niesen, we also establish a new outer bound and two new inner bounds in which it is shown to recover the centralized and decentralized strategies, previously established by Maddah-Ali and Niesen.
Finally, in terms of rate vs. cache size tradeoff, we improve the previous multiplicative gap of $72$ to $4.7$ for the average case with uniform requests.
\end{abstract}

\section{Introduction}
\begin{figure}[h!]
\footnotesize
\begin{center}
\psfrag{d1}[c]{\sf Database}
\psfrag{s1}[c]{\sf Data Server}
\psfrag{b1}[c]{\sf }
\psfrag{u1}[c]{\sf User 1}
\psfrag{u2}[c]{\sf User 2}
\psfrag{uL}[c]{\sf User $L$}
\psfrag{vd}[c]{$\vdots$}
\psfrag{c1}[c]{\sf Cache}
\psfrag{c2}[c]{\sf Cache}
\psfrag{cL}[c]{\sf Cache}
\includegraphics[width=0.45\textwidth]{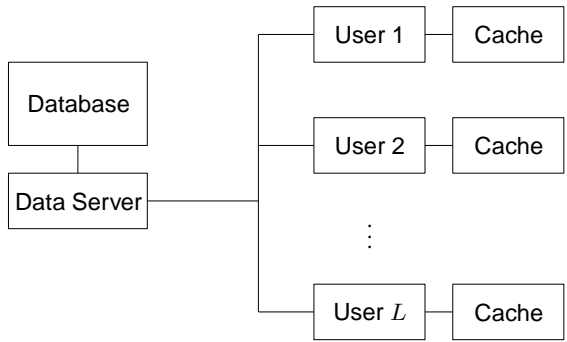}
\caption{A cache aided $L$-user file selection network.}\label{fig:fsn}
\end{center}
\end{figure}

Consider a cache-aided network that consists of a data server and $L$ users depicted in Figure~\ref{fig:fsn}. We assume that the data server has $N$ equal size files each consisting of $k$ bits, and further assume that each user is equipped with a cache of size $k\Rcache$ bits, where $\Rcache$ is the `rate' of the cache size normalized by the file length. Ideally, the data server places some description of the database during off peak hours in the users' caches such that, when the actual file requests  take place (most likely in peak hours), the total $kR_{\sf u}$  bits sent to the users to recover the individual desired files is minimized. In the considered scenario, memory is traded for peak hour bandwidth. How can such trade be made efficiently? What is the fundamental tradeoff between cache memory size and update rate?

To put the problem in perspective, we consider the following example of a {\em file selection network}\footnote{The formal definition of a file selection network is given in Section~\ref{sec:problemstatement}.}. Assume that the database has $N$ equal length files. Each file in the database consists of $k$ subfiles. Let $\mathbf{X}^{(n)}=[X_1^{(n)}, \ldots, X_k^{(n)}]$, $n\in[1:N]$ be an i.i.d. $k$-length sequence that represents the $n$th file in the database. Here, each element $X_{i}^{(n)}$, $i=1,\ldots,k$ represents the $i$th subfile of $\mathbf{X}^{(n)}$ and the collection of source vectors $(\mathbf{X}^{(1)},\ldots, \mathbf{X}^{(N)})$ represents a set of $N$ independent files in the database. Before the actual requests take place, the server caches some part of the database at each user. For each $i\in[1:k]$, we assume that each user requests a subfile from the database, namely, user $\ell\in[1:L]$ selects one subfile from $(X_i^{(1)},\ldots, X_i^{(N)})$ for each $i\in[1:k]$ from the database. The index of the file requested by user $\ell$ for $i\in[1:k]$ is represented by the random variable $Y_{\ell i}$.
For example, $Y_{\ell 1}=1, Y_{\ell 2}=4,\ldots, Y_{\ell k}=5$ corresponds to the case that decoder $\ell$ wishes to recover the sequence of subfiles $X^{(1)}_1$, $X^{(4)}_2,\ldots X^{(5)}_k$. Under this formulation, the popularity of the files (or the users' preferences) can be represented by the distribution on $Y_\ell$.

\begin{figure}[t!]
\footnotesize
\begin{center}
\psfrag{ts}[c]{Memory sharing}
\psfrag{rc}[c]{$R_{\sf c}$}
\psfrag{ru}[c]{$R_{\sf u}$}
\psfrag{qq}[c]{$\qquad\qquad$ Optimal tradeoff}
\psfrag{fx}[l]{$H(X^{(Y_1)},\ldots, X^{(Y_L)})$}
\psfrag{hx}[l]{$H(X^{(1)},\ldots, X^{(N)})$}
\psfrag{z1}[l]{$0$}
\psfrag{u1}[l]{$\mathsf{u}_L$}
\psfrag{u2}[l]{$\mathsf{u}_1$}
\includegraphics[width=0.46\textwidth]{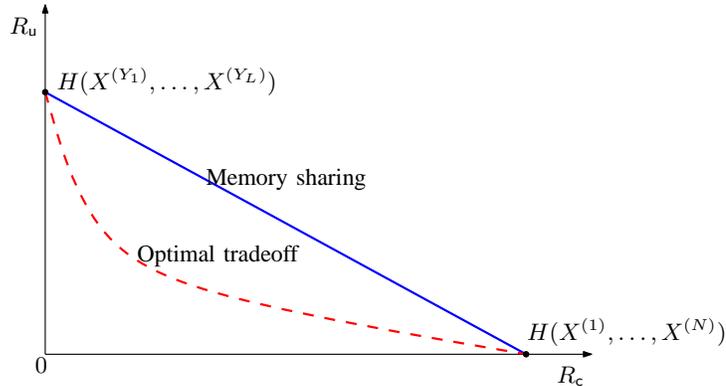}
\caption{Cache memory size vs. update delivery rate tradeoff for file selection networks. The two extreme points is achieved by either sending all the requested files or by caching the whole database. The tradeoff represented by the solid curve is attained by memory sharing between the extreme points. The non-increasing and convex optimal tradeoff curve will lie inside the memory sharing tradeoff curve.}\label{fig:tradeoff}
\end{center}
\end{figure}

Now, consider the extreme case when $\Rcache=0$. Then, by the fundamental theorem of data compression~\cite{Shannon1948}, the total number of bits required to serve all the users is $kH(X^{(Y_1)}, \ldots, X^{(Y_L)})$.  In the other extreme with $\Rcache=H(X^{(1)},\ldots, X^{(N)})$, i.e., every user has enough memory to store the whole database, the data server does not need to send anything. By {\em memory sharing} between these two extremes, i.e., we store a common fraction of the database in all the users' caches and the data server sends the remaining bits of the requested files, a straight-line tradeoff curve that connects between these two extreme points is attained; see Figure~\ref{fig:tradeoff}.
A simple improvement over this strategy is to cache the most popular files, i.e., prioritize the common cache content based on the popularity of the files. Restating the previous question: How far can we push the tradeoff curve towards the origin?

Indeed, by formulating a cache-aided network in terms of a {\em distributed source coding problem}, the authors have previously studied and characterized the fundamental limits of caching in~\cite{Wang--Lim--Gastpar2015} for single user networks (with arbitrary source and request distributions) and some two-user cache aided networks where exact solutions essentially follow from the single-user case. Moreover, in~\cite{Wang--Lim--Gastpar2015}, it was revealed that the caching problem had interesting connections to well studied information theoretic formulations, for instance, source coding with side information~\cite{Wyner--Ziv1976}, coding for computing~\cite{Orlitsky--Roche2001}, the Gray--Wyner network~\cite{Gray--Wyner1974}, the problem of successive refinement~\cite{Koshelev1985, Equitz--Cover1991}, and Wyner's common information~\cite{Wyner1975a}.

In this paper, we restrict the general assumption on the joint distribution of the source and users' requests as studied in~\cite{Wang--Lim--Gastpar2015} to the assumption that the source and users' requests are independent. This restriction (which still includes the important file section network formulation) enables a more tractable environment to study cache aided networks with {\em arbitrary number of users}; this paper is a generalization of~\cite{Wang--Lim--Gastpar2015} to the multi-user setting under the restricted distribution.

\begin{figure}[t!]
\footnotesize
\begin{center}
\psfrag{m1}[c]{$X^k$}
\psfrag{mh1}[l]{$\hat{f}_1(X^k, Y^k)$}
\psfrag{mh2}[l]{$\hat{f}_2(X^k, Y^k)$}
\psfrag{mhl}[l]{$\hat{f}_K(X^k, Y^k)$}
\psfrag{ce}[c]{Cache Enc.}
\psfrag{ue}[c]{Update Enc.}
\psfrag{d1}[c]{Decoder 1}
\psfrag{d2}[c]{Decoder 2}
\psfrag{dl}[c]{Decoder $L$}
\psfrag{rc1}[l]{$R_{\mathsf{c}1}$}
\psfrag{rc2}[l]{$R_{\mathsf{c}2}$}
\psfrag{rc3}[l]{$R_{\mathsf{c}L}$}
\psfrag{ru}[c]{$R_{\mathsf{u}}$}
\psfrag{vd}[c]{$\vdots$}
\psfrag{v1}[c]{$~Y^k$}
\psfrag{v2}[c]{$~Y^k$}
\psfrag{vl}[c]{$~Y^k$}
\psfrag{v3}[c]{$~Y^k$}
\includegraphics[width=0.55\textwidth]{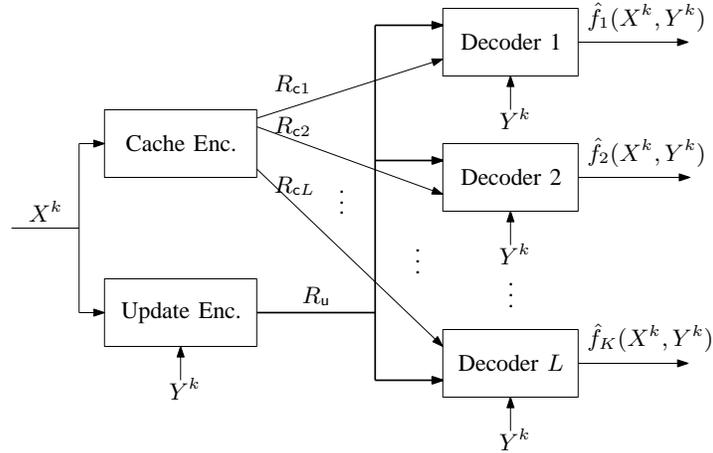}
\caption{The information theoretic $L$-user cache network (cf. Figure~\ref{fig:fsn}). The cache encoder has separate noiseless links with rate $R_{\mathsf{c}\ell}$ connected to decoder $\ell\in[1:L]$ and the update encoder has a common noiseless link to all the decoders with rate $R_{\mathsf{u}}$. The update encoder and decoders have access to the user request side information $Y^k$. }\label{fig:cache-net}
\end{center}
\end{figure}

In the next section, we first give a formal problem statement of a distributed source coding network with side information.
The network consists of two encoders, a {\em cache encoder} and an {\em update encoder} and $L$ decoders. We assume a discrete memoryless source $(X, Y)\sim p(x)p(y)$, where $X^k$ is observed at both sources, and the side information $Y^k$ is observed only at the update encoder and the $L$ decoders; see Figure~\ref{fig:cache-net}. The  objective of decoder $\ell\in[1:L]$ is to recover a {\em function} of the source and side information $f_\ell(X_i, Y_i)$, $i=1,\ldots, k$. The cache encoder has a separate link of rate $R_{\sf c \ell}$ connecting to decoder $\ell\in[1:L]$, and the update encoder is assumed to have a common link of rate $R_{\sf u}$ to all decoders.

The main motivation for studying cache aided networks in the above setup is two-fold. First, it reveals a stronger connection to distributed source coding problems which is armed with a rich set of coding theorems. With this formulation at hand and by utilizing information theoretic arguments, we provide a general outer and an inner bound for the general setup in Theorem~\ref{thm:outer-dms} and Theorem~\ref{thm:inner-dms}, respectively. Second, the general approach provides more flexibility and a unified treatment that enables extensions of these fundamental theorems to different models and assumptions. Indeed, the file selection network is a specific instance of the general distributed source coding formulation which can be represented by specifying the discrete memoryless source pair $(X, Y)$, and the functions $f_\ell(X, Y)$, $\ell\in[1:L]$ (formal statement is given in Section~\ref{sec:problemstatement}). The specialization of Theorem~\ref{thm:outer-dms} and Theorem~\ref{thm:inner-dms} to file selection networks is established in Theorem~\ref{thm:outer-FSN} for the outer bound and Theorems~\ref{thm:inner-centralized} and~\ref{thm:inner-decentralized} for the inner bound. By comparing the outer and inner bounds for uniform requests, we show that the inner bound is within a multiplicative gap of $4$ to the outer bound.
Another important aspect of this problem formulation is the flexibility that enables to extend our results to the framework of Maddah-Ali and Niesen~\cite{Maddah-Ali--Niesen2014}, i.e., when the request is constant and does not change along with the source. In particular, we provide a new outer bound (Proposition~\ref{prop:single-enc-converse}) and a new inner bound (Proposition~\ref{prop:single-enc-inner}) and show that the inner bound recovers the results~\cite[Theorem~1]{Maddah-Ali--Niesen2014} and~\cite[Theorem~2]{Maddah-Ali--Niesen2015}, but from a different path. By comparing the new outer bound and the inner bound, we improve the previous multiplicative gap of $72$ in~\cite{Maddah-Ali--Niesen2015} to $4.7$ for the average rate vs. cache size tradeoff with uniform requests, and improve the previous multiplicative gap of $12$ in~\cite{Maddah-Ali--Niesen2014} to $4.7$ for the worst case rate vs. cache size tradeoff. The extensions and statement of these results for the framework in~\cite{Maddah-Ali--Niesen2014} is given in  Section~\ref{sec:MN-extension}.

The remaining part of the paper is organized as follows. In Section~\ref{sec:converse} we collectively treat and prove the converse bounds stated throughout the paper.  In Section~\ref{sec:inner} we develop and analyze the coding strategies that establish the inner bounds. Numerical studies including some notes on the optimization of the achievable rate regions is dicussed in Section~\ref{sec:numerical}, which is followed by some concluding remarks in Section~\ref{discussion}. The lengthy proofs are deferred to the appendices.
\subsection{Previous results}
The pioneering work of Maddah-Ali and Niesen in~\cite{Maddah-Ali--Niesen2014} first demonstrated that {\em coded} caching can significantly outperform uncoded caching strategies. This important observation led to several followup works on decentralized caching~\cite{Maddah-Ali--Niesen2015}, non-uniform users requests~\cite{Niesen--Maddah-Ali2013, Zhang--Lin--Wang2015}, delay-sensitive \cite{Niesen:14d}, online \cite{Pedarsani:14}, multiple layers \cite{Karamchandani14}, request of multiple items \cite{Ji14m}, secure delivery \cite{Sengupta:15}, improved outer bounds~\cite{Tian2015, Ghasemi--Ramamoorthy2015}, caching with distortion constraints~\cite{Timo--Bidokhti--Wigger--Geiger2016}, wireless networks \cite{Ji14b, Hachem14, Jeon--Hong--Ji--Caire2015}, and improved order-optimality results~\cite{Zhang--Lin--Wang2015,Ji14z}.

\subsection{Notation}
We closely follow the notation in~\cite{El-Gamal--Kim2011}.
In particular, for a discrete random variable $X \sim p(x)$ on an alphabet $\Xc$, and for some $\e \in (0,1)$, we define the set of $\e$-typical $n$-sequences
 $x^n$ (or the typical set in short)~\cite{Orlitsky--Roche2001} as
$\aep(X) = \{ x^n : | \pi(x|x^n) - p(x) | \le \e p(x)
\text{ for all } x \in \Xc \}$, where $\pi(x|x^n)$ is the empirical pmf of $x^n$.
We use $\delta(\e) > 0$ to denote a generic function
 of $\e > 0$ that tends to zero as $\e \to 0$. A sequence of random variables is denoted by $X^k:=(X_{1},\ldots, X_{k})$. A tuple of random variables is denoted by $X(\Ac):= (X_j: j\in \Ac)$.

%
%
\section{Problem Setup and Main Results} \label{sec:problemstatement}
Let $(X, Y)$ be a pair of independent discrete memoryless sources.
A $(2^{kR_{\mathsf{c}1}},\ldots, 2^{kR_{\mathsf{c}L}}, 2^{kR_{\mathsf{u}}})$ code for the cache network consists of
\begin{itemize}
\item {\em A cache encoder} which assigns an index tuple $(m_1,\ldots, m_L)(x^k)\in[1:2^{kR_{\mathsf{c}1}}]\times\cdots\times [1:2^{kR_{\mathsf{c}L}}]$ to each sequence $x^k\in\Xc^k$,
\item {\em An update encoder} which assigns an index $m(x^k, y^k) \in[1:2^{kR_{\mathsf{u}}}]$ to each $(x^k, y^k)\in\Xc^k\times \Yc^k$, and
\item {\em $L$ decoders}, where decoder $\ell\in[1:L]$ assigns an estimate $\hat{f}_\ell(X^k, Y^k)$ to each $(m_\ell, m, y^k)$.
\end{itemize}

The performance metric is the {\em average} probability of error,
\begin{align*}
P_e^{(k)}=\P\{\hat{f}_\ell(X^k, Y^k) \neq f_\ell(X^k, Y^k) \text{ for some } \ell\in[1:L]\}.
\end{align*}
We say that a rate tuple $(R_{\mathsf{c}1},\ldots, R_{\mathsf{c}L}, R_{\mathsf{u}})$ is achievable if there exists a sequence of $(2^{kR_{\mathsf{c}1}},\ldots, 2^{kR_{\mathsf{c}L}}, 2^{kR_{\mathsf{u}}})$ codes such that $\lim_{k\to\infty}P_e^{(k)}=0$.
The optimal {\em rate--cache region} $\Rc^\star$ is the closure of the set of achievable rate tuples.
By designing efficient strategies for joint cache placement and update information processing, our goal is to characterize the fundamental tradeoff between memory size and the update bandwidth required to recover the desired contents.

Motivated by practical content delivery networks, we further specify the definition to a {\em file selection network} (FSN) setup by the following.
Let $X^k=X_1,\ldots,X_k$, $p(x^k)=\prod_{i=1}^k p(x_i)$, where each $X_i$ is an $N$-length vector
\begin{align*}
X_i=\left[X_i^{(1)},\ldots, X_i^{(N)}\right],
\end{align*}
and the components $X_i^{(n)}$, $n\in[1:N]$ are independent $\Bern(1/2)$ random variables\footnote{Since we define the rates by normalizing with respect to the source file size, assuming $\Xc_i^{(n)}$ to be binary is without loss of generality, i.e., the results remain the same if we assume $|\Xc_i^{(n)}|=q$ and $X_i^{(n)}\sim\U([1:q])$.}.
Further assume that the side information $Y^k$ is independent of $X^k$, where $Y_i$ consists of $L$ components, $Y_i=[Y_{1i},\ldots, Y_{Li}]$, $\Yc_{\ell i}=[1: N]$, $\ell\in[1:L]$, and $Y_{\ell i}$ are independent of each other.
Overall, we have the following joint distribution
\begin{align}
(X^k, Y^k) &\sim \prod_{i=1}^k p(x_{i}) p(y_{i})\nonumber\\
&= \prod_{i=1}^k\left(\prod_{n=1}^N p(x^{(n)}_{i})\prod_{\ell=1}^L p(y_{\ell i})\right). \label{eq:mr-source}
\end{align}
We assume that decoder $\ell\in[1:L]$ wishes to recover
\begin{align*}
f_\ell(X^k, Y^k)=\left[X_1^{(Y_{\ell1})}, \ldots, X_k^{(Y_{\ell k})}\right].
\end{align*}
With slight abuse of notation, we denote
$
f_\ell(X, Y)=X^{(Y_\ell)}.
$
In the sequel, we simply refer to this network as FSNs.
When we specialize our results to FSNs, we further assume a symmetric setting, i.e., we assume symmetric cache memory $R_{\mathsf{c}1}=\cdots=R_{\mathsf{c}L}=R_{\mathsf{c}}$ and
we assume that $Y_\ell$, $\ell\in[1:L]$ are independently and identically distributed, i.e., $p_Y(y)=\prod_{\ell=1}^L p_{Y_\ell}(y_\ell)$ and $p_{Y_1}=\cdots=p_{Y_L}$. For notational convenience, we denote $p_n=p_{Y_1}(n)$, $n\in[1:N]$. We assume without loss of generality that $p_1\ge p_2 \ge \cdots \ge p_N$. For some achievable rate region $\Rc$, let $\cl(\Rc)$ be its closure. When possible, we will simply express the tradeoff in terms of its {\em rate--cache tradeoff function} of $\Rc$, i.e., for some achievable rate--cache region $\Rc$,
\begin{align*}
R_{\mathsf{u}}(R_{\mathsf{c}}) = \min_{(R_{\mathsf{c}}, R_{\mathsf{u}})\in\cl(\Rc)}R_{\mathsf{u}}.
\end{align*}
Adopting from the the rate--distortion function in rate--distortion theory, the rate--cache tradeoff function for $\Rc^\star$ is simply referred to as {\em the rate--cache function} $R^\star_{\mathsf{u}}(R_{\mathsf{c}})$. Note that $R^\star_{\mathsf{u}}(R_{\mathsf{c}})$ is non-increasing and due to memory sharing (the equivalent of time sharing in distributed source coding), is convex.

We are ready to state our main results.
\subsection{Converse Bounds}
In Section~\ref{sec:converse}, we establish the following outer bound on the optimal rate--cache region.
\begin{theorem}[General lower bound]\label{thm:outer-dms}
If  a rate tuple $(R_{\mathsf{c}1},\ldots, R_{\mathsf{c}L}, R_{\mathsf{u}})$ is achievable, then
it satisfies
\begin{align*}
\sum_{\ell\in\Sc}R_{\mathsf{c}\ell} &\ge I(X;V(\Sc)), \\
R_{{\sf u}} &\ge H(F(\Sc)|V(\Sc),Y),
\end{align*}
for all $\Sc\subseteq[1:L]$ and some conditional pmf $p_{V^L|X}$, where
$F(\Sc)=\{f_\ell(X, Y): \ell\in\Sc\}$.
\end{theorem}
\smallskip

The outer bound is established by a cutset argument in which we assume that nodes in $\Sc\subseteq[1:L]$ cooperate, i.e., the decoders in $\Sc^c$ are inactive while the decoders in $\Sc$ recover $F(\Sc)$ by sharing the caches.
The proof of this theorem is given in Section~\ref{sec:converse}.

By specializing Theorem~\ref{thm:outer-dms} to FSNs, we establish the following closed-form converse bound.
\begin{theorem}[FSN lower bound]\label{thm:outer-FSN}
For FSNs with $R_{\mathsf{c}}\in[0,N]$,
\begin{align} \label{eq:WLGouter}
R^\star_{\sf u}(R_{\mathsf{c}}) \ge \max_{\ell\in[1:L]}\sum_{n=1}^N (s_n(\ell)-s_{n+1}(\ell))\left(n-\ell R_{\sf c}\right)^+,
\end{align}
where $(x)^+=\max(x,0)$, $s_{N+1}(\ell)=0$ and $s_n(\ell) = 1 - (1-p_n)^\ell$, $n\in[1:N]$.
\end{theorem}
\smallskip

By setting $p_{Y_\ell} = \text{Unif}([1:N])$, $\ell\in[1:L]$ in Theorem~\ref{thm:outer-FSN}, we have the following simplified converse bound for the uniform case.
\begin{corollary}[FSN lower bound for uniform requests] \label{cor:outer-uni}
For FSNs with uniform requests and $R_{\mathsf{c}}\in[0,N]$,
\begin{align} \label{eq:outer_uni}
R_{\sf u}^\star(R_{\sf c}) &\ge \max_{\ell\in[1:L]} (1-(1-1/N)^\ell )(N-\ell R_{\sf c})^+.
\end{align}
\end{corollary}
\smallskip

\subsection{Inner Bounds}
In Section~\ref{sec:inner}, we establish the following inner bounds on the optimal rate--cache region.
The general coding scheme and its specialization for FSNs with centralized and decentralized caching constitute the key contributions for achievability.

\begin{theorem}[General Inner bound]\label{thm:inner-dms}
A rate tuple $(R_{\mathsf{c}1},\ldots, R_{\mathsf{c}L}, R_{\mathsf{u}})$ is achievable if
\begin{align*}
R_{\mathsf{c}\ell} &> I(V_\ell; X|Q), \quad \ell \in[1:L] \\
R_\mathsf{u} &> \sum_{\substack{\Sc\subseteq[1:L]}}\max_{\ell\in\Sc} I(U_{\Sc}; X|V_\ell, Y, Q)
\end{align*}
for some $p(q)\prod_{\ell=1}^L p(v_\ell|x, q)\prod_{\Sc\subseteq[1:L]}p(u_{\Sc}|x, y, q)$ such that
\begin{align}
H(f_\ell(X, Y)|(U_\Sc:  \Sc\subseteq[1:L], \ell\in\Sc), V_\ell, Y, Q)=0, \label{eq:ergodic-cond}
\end{align}
for all $\ell\in[1:L]$.
\end{theorem}

For FSNs, Theorem~\ref{thm:inner-dms} can be specialized to the following Theorems. 
By a specific choice of auxiliary random variables given in Section~\ref{sec:centralized} we establish the first FSN inner bound in the following theorem.

\begin{theorem}[Centralized caching for FSNs]\label{thm:inner-centralized}
For FSNs and $R_{\mathsf{c}}=0,\frac{1}{L},\frac{2}{L},\ldots,N$,
\begin{align}
R^\star_{\mathsf{u}} (R_{\mathsf{c}})&\le \sum_{n=1}^N\sum_{j=1}^{L-r_n}\frac{j}{j+r_n}  \binom{L-r_n}{j} p_n^{j}(1-p_n)^{L-r_n-j}, \label{eq:inner-centralized}
\end{align}
for $r_n\in[0:L]$, $n\in[1:N]$ such that $\sum_{n=1}^Nr_n=LR_{\mathsf{c}}$.
\end{theorem}
The proof of this theorem is given in Subsection~\ref{sec:centralized}.
\smallskip
\begin{remark}
The achievable rate--cache tradeoff in \eqref{eq:inner-centralized} is defined for $R_{\mathsf{c}}=0,\frac{1}{L},\frac{2}{L},\ldots,N$ such that $\sum_{n=1}^Nr_n=LR_{\mathsf{c}}$ for some $r_n\in[0:L]$, $n\in[1:N]$. The rest of the points in $\Rcache\in[0, N]$ are obtained by memory-sharing between these discrete points resulting in a piece-wise linear tradeoff function.
\end{remark}

For the case with uniform requests, we establish the following corollary.
\begin{corollary}[Centralized caching for uniform requests]\label{cor:centralized-unif}
For FSNs with uniform requests and $R_{\mathsf{c}}=\frac{N}{L},\frac{2N}{L},\frac{3N}{L},\ldots,N$,
\begin{align*}
R^\star_{\mathsf{u}} (R_{\mathsf{c}})&\le N\E\left[\frac{Z}{Z+r}\right],
\end{align*}
for $r\in[1:L]$ such that $r=LR_{\mathsf{c}}/N$, where $Z\sim \text{Binom}(L-r,1/N)$. Moreover, for $R_{\mathsf{c}}=0$,
\begin{align*}
R^\star_{\mathsf{u}} (R_{\mathsf{c}})&= N(1-(1-1/N)^L).
\end{align*}
\end{corollary}
\smallskip

By a different choice of the auxiliary random variables given in Section~\ref{sec:decentralized}, Theorem~\ref{thm:inner-dms} can also be specialized to the following inner bound for FSNs.
\begin{theorem}[Decentralized caching inner bound]\label{thm:inner-decentralized}
For FSNs with $R_{\sf c}\in[0,N]$,
\begin{align}
R^\star_{\sf u}(R_{\sf c}) &\le \sum_{n=1}^N\frac{p_n(1-r_n)}{1-\alpha_n} \left[1-\alpha_n^L\right], \label{eq:inner-decentralized}
\end{align}
for $r_n\in[0,1]$ such that $\sum_{n=1}^N r_n =R_{\sf c}$, where $\alpha_n=(1-p_n)(1-r_n)$.
\end{theorem}
The proof of this theorem is given in Subsection~\ref{sec:decentralized}.
\smallskip
\begin{remark}
We call the strategy that attains Theorem~\ref{thm:inner-decentralized} `decentralized' due to the additional feature that, if $r_n$, $n\in[1:N]$ is chosen only based on the file popularity distribution, then the {\em cache encoder} is decentralized. Following the convention of~\cite{Maddah-Ali--Niesen2015}, we say that a cache encoder is decentralized if $(m_1,\ldots, m_L)(x^k)=(m_1(x^k),\ldots, m_L(x^k))$, i.e., the cache encoder mapping for user $\ell$ does not depend on the mappings of the other users messages.
\end{remark}

By further assuming uniform requests, we simplify Theorem~\ref{thm:inner-decentralized} to the following corollary.
\begin{corollary}[Decentralized caching for uniform requests]\label{cor:decentralized-unif}
For FSNs with uniform requests and $R_{\sf c}\in[0,N]$,
\begin{align} \label{eq:dec-unif}
R^\star_{\sf u}(R_{\sf c}) &\le \frac{N-R_{\sf c}}{1+R_{\sf c}(1-1/N)} \left[1-\left(\left(1-\frac{1}{N}\right)\left(1-\frac{R_{\sf c}}{N}\right)\right)^L\right].
\end{align}
\end{corollary}
\smallskip
Denote by $\bar{R}_{\sf u\tdash dc}(R_{\mathsf{c}})$ the right hand side of~\eqref{eq:dec-unif}.
The following theorem provides a universal (in $N$ and $L$) performance guarantee of the decentralized caching strategy in terms of a multiplicative gap from the optimal tradeoff for uniform requests.
\begin{theorem}[Multiplicative gap] \label{thm:gap-ergodic}
For the case with $p_{Y_\ell}=\U([1:N])$ and $\Rcache\in[0,N)$, it holds that
\begin{align*}
\frac{\bar{R}_{\sf u\tdash dc}(R_{\mathsf{c}})}{{R}^\star_{\mathsf{u}}(R_{\mathsf{c}})} \le 4.
\end{align*}
\end{theorem}
The proof of this theorem is given in Appendix~\ref{app:gap-erogdic}.

The centralized strategy is optimal for some high-cache regime stated in the following corollary.
\begin{corollary}\label{cor:optimal-regime}
For FSNs with arbitrary request distributions and $R_{\sf c}\in\left[N-1/L, N\right]$,
\begin{align}
R^\star_{\mathsf{u}} (\Rcache) = p_N(N-\Rcache). \label{eq:opt-regime}
\end{align}
Moreover, for uniform requests, \eqref{eq:opt-regime} holds for $R_{\sf c}\in\left[N-N/L, N\right]$.
\end{corollary}
The proof is given in Appendix~\ref{app:optimal-regime}.

\section{New Results for the Framework of Maddah-Ali and Niesen}\label{sec:MN-extension}
The framework studied in this paper was motivated by the pioneering work of Maddah-Ali and Niesen~\cite{Maddah-Ali--Niesen2014} on coded caching. The main difference is in the approach we take for tackling the problem, that is, we take an information theoretic approach by viewing the problem as a distributed source coding problem. In this section, we extend the results of the previous section to the framework of~\cite{Maddah-Ali--Niesen2014}.

We begin by formulating an extension of our problem setup in which the request information $Y$ changes only every $T$ source symbols $X^T$ and $f_\ell(X^T, Y)=(f_{\ell,1}(X_1, Y),\ldots,f_{\ell,T}(X_T, Y))$. We refer to this model as the {\em static request model}\footnote{A general discussion on the comparison of the models can be found in~\cite[Section VI]{Wang--Lim--Gastpar2015}.}.
By treating each block as a ``super-symbol'' and coding over $kT$ symbols and applying Theorem~\ref{thm:inner-dms}, a rate tuple $(R_{\mathsf{c}1},\ldots, R_{\mathsf{c}L},R_{\sf u})$ is achievable if
\begin{align*}
TR_{\mathsf{c}\ell} &> I(\Vt_\ell; X^T|Q), \quad \ell \in[1:L] \\
TR_\mathsf{u} &> \sum_{\substack{\Sc\subseteq[1:L]}}\max_{\ell\in\Sc} I(\Ut_{\Sc}; X^T|\Vt_\ell, Y, Q)
\end{align*}
for some $p(q)\prod_{\ell=1}^L p(\vt_\ell|x^T, q)\prod_{\Sc\subseteq[1:L]}p(\ut_{\Sc}|x^T, y, q)$ such that
\begin{align}
H(f_\ell(X^T, Y)|(\Ut_\Sc:  \Sc\subseteq[1:L], \ell\in\Sc), \Vt_\ell, Y, Q)=0, \quad \ell\in[1:L]. \label{eq:comp-cond}
\end{align}
By choosing $\Vt_\ell=V^T_{\ell}$ and $\Ut_\Sc=U^T_{\Sc}$ such that $p(\vt_\ell|x^T)=\prod_{i=1}^T p_{V|X, Q}(v_{\ell,i}|x_i,q)$ and $p(\ut_\Sc|x^T)=\prod_{i=1}^T p_{U_\Sc|X, Y, Q}(u_{\Sc,i}|x_i,y,q)$, we can conclude that the exact expression in Theorem~\ref{thm:inner-dms} is also achievable for the static request model, if we allow {\em encoding over multiple blocks}. In this sense, the corresponding rate region provides an {\em ergodic} achievable rate--cache tradeoff $R_{\sf u}(R_{\sf c})$.

One the other hand, consider the case when the encoders are restricted to encode over each block separately.\footnote{In the case for encoding over multiple blocks, the total number of blocks is assumed to be sufficiently large. On the other hand, for coding within a single block, the number of symbols in a block is assumed to be sufficiently large. The fitness of the two models for practical networks depends on the underlined assumption of how frequent the requests change compared to the file size.} Naturally, we define a rate--cache region for this case as a set of achievable rate tuples $(2^{nR_{\mathsf{c}}}, (2^{nR_{\mathsf{u}}(y)}: y\in\Yc))$, where $R_{\sf u}(y)$ is the update rate when the request side information is $y$. The corresponding rate--cache tradeoff function $R_{\sf u}(R_{\sf c}, y)$ is thus defined for each $y\in\Yc$.
Depending on the application criteria, we can further formulate the problem statement in the following ways. 
Based on a rate region $\Rc$ for the static request single block encoding setup, the update rate tuples can be projected to:
\begin{enumerate}
\item the {\em worst case update rate} or {\em compound rate}
\begin{align*}
R_{\mathsf{u\tdash wc}}(R_{\sf c})=\max_{y\in\Yc} R_\mathsf{u}(R_{\sf c},y),
\end{align*}
\item the {\em average rate}
\begin{align*}
R_{\mathsf{u \tdash ave}}(R_{\sf c})=\E_Y[R_\mathsf{u}(R_{\sf c}, Y)].
\end{align*}
\end{enumerate}
We denote by $R^\star_{\mathsf{u\tdash wc}}(R_{\sf c})$ and $R^\star_{\mathsf{u \tdash ave}}(R_{\sf c})$ the optimal worst case rate--cache function and the optimal average rate--cache function, respectively.
For static request model, define a FSN by
\begin{align}
(X^k, Y) &\sim \prod_{i=1}^k p(x_{i}) p(y)\nonumber\\
&= \left(\prod_{i=1}^k\prod_{n=1}^N p(x^{(n)}_{i})\right)\prod_{\ell=1}^L p(y_{\ell}), \label{eq:sr-source}
\end{align}
and assume that decoder $\ell\in[1:L]$ wishes to recover
\begin{align*}
f_\ell(X^k, Y)=\left[X_1^{(Y_{\ell})}, \ldots, X_k^{(Y_{\ell})}\right].
\end{align*}
For the static request FSN, the work of Maddah-Ali and Niesen in~\cite{Maddah-Ali--Niesen2014} studies the tradeoff between $R_{\mathsf{c}}$ and the worst case rate in~\cite{Maddah-Ali--Niesen2015, Maddah-Ali--Niesen2014}, and the tradeoff between $\Rcache$ and the average rate in~\cite{Niesen--Maddah-Ali2013}.

In the following, we discuss some extensions of our results to the static request model with single block encoding.

\begin{proposition}[Converse Bound]\label{prop:single-enc-converse}
For the static request model with single block encoding, if  a rate tuple $(R_{\mathsf{c}1},\ldots, R_{\mathsf{c}L}, R_{\mathsf{u}}(y), y\in\Yc)$ is achievable, then it satisfies
\begin{align}
\sum_{\ell\in\Sc}R_{{\sf c}\ell} &\ge I(X;V(\Sc)), \nonumber\\
R_{{\sf u}}(y) &\ge H(F(\Sc)|V(\Sc),Y=y), \quad y\in\Yc, \label{eq:sb-Ru}
\end{align}
for all $\Sc\subseteq[1:L]$ and some conditional pmf $p_{V^L|X}$. Moreover, if an average rate is achievable, it satisfies
\begin{align}
\E_Y[R_{{\sf u}}(Y)] &\ge \E_Y[H(F(\Sc)|V,Y=y)]\\
&= H(F(\Sc)|V,Y), \label{eq:ave-converse}
\end{align}
and if a worst case rate is achievable, it satisfies
\begin{align*}
\max_{y\in\Yc} R_{{\sf u}}(y) &\ge \max_{y\in\Yc}  H(F(\Sc)|V,Y=y).
\end{align*}
\end{proposition}
The proof of this proposition is given in Section~\ref{sec:converse}.

\begin{remark}
For the static request FSN with single block encoding, due to~\eqref{eq:ave-converse}, the outer bound for the average rate--cache region in Proposition~\ref{prop:single-enc-converse} has the same expression as Theorem~\ref{thm:outer-dms}. As a consequence, Theorem~\ref{thm:outer-FSN} and Corollary~\ref{cor:outer-uni} also apply to $R^\star_{\sf u\tdash ave}(\Rcache)$. Consequently, Theorem~\ref{thm:outer-FSN} and Corollary~\ref{cor:outer-uni} also apply to $R^\star_{\sf u\tdash wc}(\Rcache)$ since $R^\star_{\sf u\tdash wc}(\Rcache) \ge R^\star_{\sf u\tdash ave}(\Rcache)$.
\end{remark}

On the other hand, Theorem~\ref{thm:inner-dms} can be extended to following proposition for the single block encoding case.
\begin{proposition}[Inner Bound]\label{prop:single-enc-inner}
For the static request model with single block encoding, a rate tuple $(R_{\mathsf{c}1},\ldots, R_{\mathsf{c}L}, R_{\mathsf{u}}(y), y\in\Yc)$ is achievable if,
\begin{align}
R_{\mathsf{c}\ell} &> I(V_\ell; X|Q), \quad \ell \in[1:L] \\
R_\mathsf{u}(y) &> \sum_{\substack{\Sc\subseteq[1:L]}}\max_{\ell\in\Sc} I(U_{\Sc}; X|V_\ell, Y=y, Q), \quad y\in\Yc, \label{eq:sb-ry-inner}
\end{align}
for some $p(q)\prod_{\ell=1}^L p(v_\ell|x, q)\prod_{\Sc\subseteq[1:L]}p(u_{\Sc}|x, y, q)$ such that
\begin{align}
\max_{y\in\Yc}H(f_\ell(X, Y)|(U_\Sc:  \Sc\subseteq[1:L], \ell\in\Sc), V_\ell, Y=y, Q)=0, \quad \ell\in[1:L]. \label{eq:comp-inner-cond}
\end{align}
Moreover, an average rate $R_{\sf u\tdash ave}=\E_Y[R_{{\sf u}}(Y)]$ is achievable if,
\begin{align}
R_{\sf u\tdash ave} &> \sum_{y\in\Yc}p_Y(y)\left[\sum_{\substack{\Sc\subseteq[1:L]}}\max_{\ell\in\Sc} I(U_{\Sc}; X|V_\ell, Y=y, Q)\right] \label{eq:ave-rate}
\end{align}
and the worst case rate $R_{\sf u\tdash wc}=\max_{y\in\Yc} R_{{\sf u}}(y)$ is achievable if
\begin{align*}
R_{\sf u\tdash wc} &> \max_{y\in\Yc}  H(F(\Sc)|V,Y=y).
\end{align*}
\end{proposition}
The proof of this proposition is given in Section~\ref{sec:inner}.

\begin{remark}
Although the source--request pair $(X^k, Y^k)$ for the model in Section~\ref{sec:problemstatement}, and the source--request pair $(X^k, Y)$ for the static request model are different, the converse and achievability results for both models are evaluated under the same form of single-letter random variables $(X, Y)\sim p_{X}(x)p_{Y}(y)$.
Accordingly, if we choose a joint distribution in Proposition~\ref{prop:single-enc-inner} that results in
\begin{align*}
I(U_{\Sc}; X|V_\ell, Y=y, Q)=I(U_{\Sc}; X|V_{\ell'}, Y=y, Q),
\end{align*}
for $\Sc\subseteq[1:L]$, $\ell, \ell'\in \Sc$, $\ell\neq\ell'$, we can get rid of the maximum in equation \eqref{eq:ave-rate}.
Under such distributions the rate--cache region in Theorem~\ref{thm:inner-dms} and the average rate--cache region in Proposition~\ref{prop:single-enc-inner} are equal.
\end{remark}

By specializing Proposition~\ref{prop:single-enc-inner}, we establish a centralized rate--cache tradeoff for the static request single block encoding FSN stated in the following theorem.
\begin{theorem}[Centralized inner bound for static request] \label{thm:sbe-centralized}
Consider the static request single block encoding FSN. For $R_{\mathsf{c}}=0,\frac{1}{L},\frac{2}{L},\ldots,N$, a rate tuple $(\Rcache, (R_{\sf u}(y): y\in\Yc))$ is achievable if
\begin{align}
R_{\mathsf{u}}(\Rcache,y)&> \sum_{\Sc:|\Sc|>0}\left(1- \prod_{\ell\in\Sc}\mathbb{1}\{r_{y_\ell} \neq |\Sc| -1\}\right) \frac{1}{\binom{L}{|\Sc|-1}}, \quad y\in\Yc, \label{eq:single-enc-cent}
\end{align}
for $r_n\in[0:L]$, $n\in[1:N]$ such that $\sum_{n=1}^Nr_n=LR_{\mathsf{c}}$.
Moreover, an average rate $R_{\sf u\tdash ave}(\Rcache)=\E_Y[R_{{\sf u}}(\Rcache, Y)]$ is achievable if,
\begin{align*}
R_{\sf u\tdash ave}(\Rcache)&> \sum_{j=0}^{L-1}\frac{(L-j)}{j+1}\left(1- \left(1-\alpha_j\right)^{j+1}\right),
\end{align*}
where $\alpha_j=\sum_{n=1}^N\mathbb{1}\{r_{n} = j\}p_n$.
\end{theorem}
\smallskip
\begin{remark}\label{rmk:recover-cent}
By choosing $r_n=L\Rcache/N$ in \eqref{eq:single-enc-cent}, a worst case rate--cache tradeoff
$R_{\sf u\tdash wc}(\Rcache)$ is achievable if
\begin{align}
R_{\sf u\tdash wc}(\Rcache) > \frac{L-L\Rcache/N}{1+L\Rcache/N}.
\end{align}
This recovers the result of~\cite[Theorem~1]{Maddah-Ali--Niesen2014}. In this sense, Theorem~\ref{thm:sbe-centralized} generalizes the strategy of~\cite[Theorem~1]{Maddah-Ali--Niesen2014} to the average rate--cache tradeoffs with arbitrary request distributions. The underlined strategy that establishes the theorem is based on distributed source coding techniques instead of the explicit network coding strategy in~\cite{Maddah-Ali--Niesen2014}.
Potentially, the choice of auxiliary random variables used in the proof of Theorem~\ref{thm:inner-centralized} can improve the inner bound presented in Theorem~\ref{thm:sbe-centralized} which is based on a simpler (but easier to evaluate) choice. We refer to Appendix~\ref{app:sbe-centralized} for the explicit choice of auxiliary random variables and the proof of Theorem~\ref{thm:sbe-centralized}.
\end{remark}

Similarly, by specializing Proposition~\ref{prop:single-enc-inner}, we establish a decentralized rate--cache tradeoff for the static request single block encoding FSN stated in the following theorem.
\begin{theorem}[Decentralized inner bound for static request]\label{thm:sbe-decentralized}
Consider the static request single block encoding FSNs. For $R_{\mathsf{c}}\in[0,N]$, a rate tuple $(\Rcache, (R_{\sf u}(y): y\in\Yc))$ is achievable if
\begin{align}\label{eq:sbe-decentralized}
R_{\mathsf{u}}(\Rcache,y)&> \sum_{j=1}^L \sum_{\mathcal{S}:|\mathcal{S}|=j} \max_{\ell\in\mathcal{S}} r_{y_\ell}^{j-1}(1-r_{y_\ell})^{L-j+1},
\end{align}
for $r_n\in[0,1]$, $n\in[1:N]$ such that $\sum_{n=1}^Nr_n=R_{\mathsf{c}}$.
\end{theorem}
\begin{remark}\label{rmk:recover-decent}
By choosing $r_n=\Rcache/N$, in~\eqref{eq:sbe-decentralized} a worst case rate--cache tradeoff $R_{\sf u\tdash wc}(\Rcache)$ is achievable if
\begin{align}
R_{\sf u\tdash wc}(\Rcache) > \frac{\left( N-\Rcache\right)}{\Rcache} (1-(1-\Rcache/N)^L),\label{eq:MN-dec}
\end{align}
which recovers the result of~\cite[Theorem~1]{Maddah-Ali--Niesen2015} for decentralized caching.
\end{remark}

Denote the right hand side of~\eqref{eq:MN-dec} by $R_{\sf MN}(R_{\sf c})$.
Note that $R_{\sf MN}(R_{\sf c})$ is not convex. Thus, by memory sharing among the achievable points, the rate--cache tradeoff can be improved. We denote by $\breve{R}_{\sf MN}(R_{\sf c})$ the corresponding convexified bound.
By comparing $\breve{R}_{\sf MN}(R_{\sf c})$ with Theorem~\ref{thm:outer-FSN} we have the following theorem.
\smallskip

\begin{theorem}[Multiplicative gap for static request single block encoding]\label{thm:sb-gap}
For the static request single block encoding FSN with $p_{Y_\ell}=\U([1:N])$ and $\Rcache\in[0,N)$,
\begin{align}
\frac{\breve{R}_{\sf MN}(R_{\sf c})}{{R}^\star_{\mathsf{u\tdash ave}}(R_{\mathsf{c}})} \le 4.7. \label{eq:sb-gap}
\end{align}
\end{theorem}
\smallskip

\begin{remark}\label{rmk:gap}
We remark that Theorem~\ref{thm:sb-gap} implies that \eqref{eq:sb-gap} also holds when $R^\star_{\mathsf{ave \tdash u}}(R_{\mathsf{c}})$ is exchanged with $R^\star_{\mathsf{wc \tdash u}}(R_{\mathsf{c}})$, i.e., the worst case rate--cache function, since it is lower bounded by the average rate--cache function $R^\star_{\mathsf{ave\tdash  u}}(R_{\mathsf{c}})$.
\end{remark}

The above theorem improves upon the multiplicative gap of $72$ in~\cite{Maddah-Ali--Niesen2015}. Furthermore,
for the worst case, in light of Remark~\ref{rmk:gap}, we improve the previous gap of $12$ in~\cite{Maddah-Ali--Niesen2014}. The proof of this theorem is given in Appendix~\ref{app:sb-gap}.

\begin{remark}
In an independent work~\cite{Ghasemi--Ramamoorthy2015}, the authors introduce a lower bound specifically for the worst-case that attains a multiplicative gap of 4. Compared to the lower bound in~\cite{Ghasemi--Ramamoorthy2015}, our lower bound applies to arbitrary request distributions.
\end{remark}

\begin{figure}[h!]
\begin{center}
\psfrag{a1}[t]{$R_{\mathsf{c}}$}
\psfrag{b1}[b]{$R_{\mathsf{u}}$}
\hspace{6pt}\includegraphics[width=0.5\textwidth]{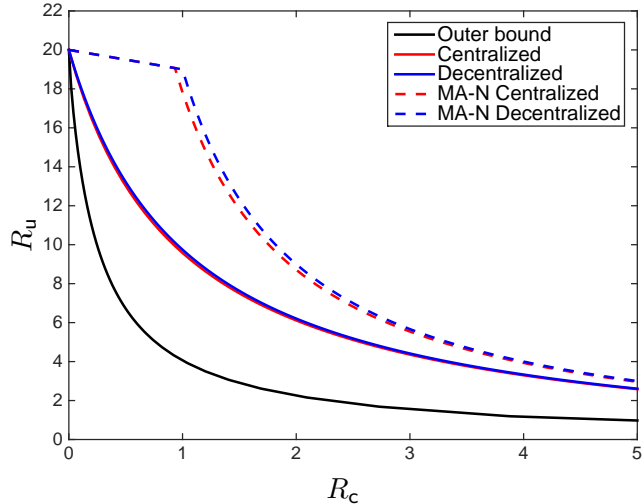}
\caption{The rate--cache tradeoff for the centralized and decentralized schemes, for $N=20$, $L=300$. The solid curves are the `ergodic' rate--cache tradeoff curves in Corollary~\ref{cor:centralized-unif} and Corollary~\ref{cor:decentralized-unif} with uniform requests, and the dashed curves are the `compound' rate--cache tradeoff in Maddah-Ali and Niesen~\cite{Maddah-Ali--Niesen2014, Maddah-Ali--Niesen2015}.}
\label{fig:compare}
\end{center}
\end{figure}

In Figure~\ref{fig:compare} we plot the performance of the `ergodic' rate--cache tradeoff curves in Corollary~\ref{cor:centralized-unif} and Corollary~\ref{cor:decentralized-unif} for uniform requests, and the `compound' rate--cache tradeoffs in Maddah-Ali and Niesen~\cite{Maddah-Ali--Niesen2014, Maddah-Ali--Niesen2015}.
For both ergodic and compound settings, the centralized strategies uniformly perform better than their respective decentralized strategies.

\section{Proof of Converse Bounds} \label{sec:converse}
In this section, we present the proof of Theorem~\ref{thm:outer-dms}, Proposition~\ref{prop:single-enc-converse}, and Theorem~\ref{thm:outer-FSN}. We begin with the proof of Theorem~\ref{thm:outer-dms}.

Consider any subset $\Sc\subseteq [1:L]$. Denote $V_{\ell i} = (M_\ell,X^{i-1})$, $\ell\in[1:L]$, $i\in[1:k]$. Since $X$ and $Y$ are independent by assumption, the Markov chain $(V_{i,1},\ldots,V_{i,L})\to X_i \to Y_i$ holds for all $i\in[1:k]$.
Then, since $H(M_\ell) \le kR_{\mathsf{c}\ell}$ for all $\ell\in[1:L]$, we have
\begin{align*}
k \sum_{\ell\in\Sc}R_{\mathsf{c}\ell} &\ge \sum_{\ell\in \Sc} H(M_\ell) \\
&\ge H(M(\mathcal{S})) \\
&= I(X^k;M(\mathcal{S})) \\
&= \sum_{i=1}^k I(X_i;M(\mathcal{S})|X^{i-1}) \\
&= \sum_{i=1}^k I(X_i;M(\mathcal{S}), X^{i-1}) \\
&= \sum_{i=1}^k I(X_i;V_{i}(\Sc)).
\end{align*}
Recall $F_\ell=f_\ell(X, Y)$. Then, we have
\begin{align*}
k R_\mathsf{u} &\ge H(M| Y^k) \\
&\ge H(M| M(\Sc), Y^k) \\
&= H(F^k(\Sc), M|M(\Sc), Y^k) - H(F^k(\Sc)|M, M(\Sc), Y^k) \\
&\stackrel{(a)}{\ge} H(F^k(\Sc)|M(\Sc), Y^k) - k\epsilon_k \\
&=\sum_{i=1}^k H(F_i(\Sc)|F^{i-1}(\Sc), M(\Sc), Y^k) - k\epsilon_k \\
&\ge \sum_{i=1}^k H(F_i(\Sc)|X^{i-1}(\Sc), M(\Sc), Y^k) - k\epsilon_k \\
&= \sum_{i=1}^k H(F_i(\Sc)|X^{i-1}(\Sc), M(\Sc), Y_i) - k\epsilon_k \\
&= \sum_{i=1}^k H(F_i(\Sc)|V_i(\Sc), Y_i) - k\epsilon_k,
\end{align*}
where $(a)$ follows from the data processing inequality and Fano's inequality, and $\epsilon_k$ tends to zero as $k\to \infty$. The rest of the proof follows from the standard time sharing argument and then letting $k\to\infty$.

Thus, we have that
\begin{align*}
R_{\sf u}^\star(R_{\sf c}) &\ge \min \max_{\mathcal{S}\subseteq [1:L]} H(F(\Sc)|V(\Sc), Y),
\end{align*}
where the minimum is over all conditional pmfs $p_{V^L|X}$ such that $V^L\to X \to Y$ form a Markov chain and
\begin{align*}
I(X;V(\Sc)) &\le \sum_{\ell\in\Sc}R_{{\sf c}\ell}, \quad \forall \Sc\subseteq[1:L].
\end{align*}
This concludes the proof of Theorem~\ref{thm:outer-dms}.

At this point, extending the proof to Proposition~\ref{prop:single-enc-converse} requires only minor changes which we highlight in the following.
For the static request model with single block encoding, the proof steps for the bound on $\Rcache$ remains the same since the cache encoder does not utilize the information of $Y$ in both cases. For the bounds on the update rate $R_{\sf u}$, the difference is that in the static request model with single block encoding, we have multiple messages $M_y$ for each $y\in\Yc$. Thus, we can redo the steps for the bounding $R_{\sf u}$ with $M_y \in[1:2^{kR_{\sf u}(y)}]$ assuming $Y=y$ which gives the condition~\eqref{eq:sb-Ru}.

Next, we prove Theorem~\ref{thm:outer-FSN}.
First, we restrict attention to the case of i.i.d. requests, i.e., $p_{Y}(y) = \prod_{\ell=1}^L p_{Y_\ell}(y_\ell)$ and $p_{Y_1}=\cdots=p_{Y_L}$. Further specializing to FSNs, we obtain a closed-form bound on $R^\star_{\sf u}(R_{\sf c})$ by switching between the $\min$ and $\max$ (and thus relaxing the bound), i.e., we have that for $\Sc\subseteq [1:L]$,
\begin{align*}
R_{\sf u}^\star(R_{\sf c}) &\ge \min_{p_{V(\Sc)|X}} H(F(\Sc)|V(\Sc), Y),
\end{align*}
such that
\begin{align*}
I(X;V(\Sc)) &\le \sum_{\ell\in\Sc}R_{{\sf c}\ell}.
\end{align*}
For $\Sc\subseteq[1:L]$ and $n\in[1:N]$, we denote
\begin{align} \label{eq:sn}
s_n(\Sc) &:= \P\{n\in Y(\Sc)\} = \sum_{\substack{ y: n\in y(\Sc)}} p_{Y}(y).
\end{align}
For simplicity, we will use the short hand notation $s_n=s_n(\Sc)$ while keeping in mind that $s_n$ depends on $\Sc$.
Without loss of generality, we assume that $s_1\ge s_2 \ge \cdots \ge s_N$.
Suppose that $(R_{\sf c},R_{\sf u})\in\mathcal{R}^\star$. Then, there exists a conditional pmf $p_{V(\Sc)|X}$ such that $\sum_{\ell\in\Sc} R_{\sf c \ell} \ge I(X;V(\Sc))=:r$ and $R_{\sf u} \ge H(F(\Sc)|V(\Sc),Y)$.
For $n\in[1:N]$, we have
\begin{align}
\sum_{\ell\in\Sc} R_{\sf c \ell} \ge r &=  I(X;V(\Sc)) \nonumber \\
&\ge I(X^{([1:n])};V(\Sc)|X^{([n+1:N])}) \nonumber \\
\label{eq:cond_subsetind}
&= H(X^{([1:n])}) - H(X^{(n)}|V(\Sc),X^{([n+1:N])}) - H(X^{([1:n-1])}|V(\Sc),X^{([n:N])}).
\end{align}
Now we show that $R_{\sf u}$ can be lower bounded as in \eqref{eq:WLGouter}.
First, we have
\begin{align*}
R_{\sf u} &\ge H(F(\Sc)|V(\Sc),Y) \\
&= \sum_{y} p_{Y}(y) H((X^{(y_\ell)}, \ell\in\Sc)|V(\Sc)) \\
&\ge \sum_{n=1}^N s_n H(X^{(n)}|V(\Sc),X^{([n+1:N])}),
\end{align*}
where the last inequality follows by recursively applying
\begin{IEEEeqnarray*} {ll}
& \sum_{y} p_Y(y) H((X^{(y_\ell)}, \ell\in\Sc)|V(\Sc),X^{([n+1:N])}) \\
&\ge s_nH(X^{(n)}|V(\Sc),X^{([n+1:N])}) \\
&\quad + \sum_{y} p_Y(y) H((X^{(y_\ell)}, \ell\in\Sc)|V(\Sc),X^{(n)}, X^{([n+1:N])}),
\end{IEEEeqnarray*}
in the order $N,N-1,\cdots,1$.
Next, $R_{\sf u}$ can be further lower bounded as
\begin{align*}
R_{\sf u} &\ge \sum_{n=1}^N s_n H(X^{(n)}|V(\Sc),X^{([n+1:N])}) \\
&= s_N H(X^{(N)}|V(\Sc)) + \sum_{n=1}^{N-1} s_n H(X^{(n)}|V(\Sc),X^{([n+1:N])}) \\
&\stackrel{(a)}{\ge} s_N \left(H(X^{([1:N])})-r-H(X^{([1:N-1])}|V(\Sc),X^{(N)})\right)^+ + \sum_{n=1}^{N-1} s_n H(X^{(n)}|V(\Sc),X^{([n+1:N])}) \\
&\stackrel{(b)}{\ge} s_N \left(H(X^{([1:N])})-r\right)^+ - s_NH(X^{([1:N-1])}|V(\Sc),X^{(N)}) + \sum_{n=1}^{N-1} s_n H(X^{(n)}|V(\Sc),X^{([n+1:N])}) \\
&= s_N \left(H(X^{([1:N])})-r\right)^+ + \sum_{n=1}^{N-1} (s_n-s_N) H(X^{(n)}|V(\Sc),X^{([n+1:N])}) \\
&= s_N \left(H(X^{([1:N])})-r\right)^+ + (s_{N-1}-s_N) H(X^{(N-1)}|V(\Sc),X^{(N)}) \\
&\quad + \sum_{n=1}^{N-2} (s_n-s_N) H(X^{(n)}|V(\Sc),X^{([n+1:N])}) \\
&\stackrel{(c)}{\ge} s_N \left(H(X^{([1:N])})-r\right)^+ + (s_{N-1}-s_N) \left(H(X^{([1:N-1])})-r - H(X^{([N-2])}|V(\Sc),X^{([N-1:N])})\right)^+ \\
&\quad + \sum_{n=1}^{N-2} (s_n-s_N) H(X^{(n)}|V(\Sc),X^{([n+1:N])}) \\
&\stackrel{(d)}{\ge} s_N \left(H(X^{([1:N])})-r\right)^+ + (s_{N-1}-s_N) \left(H(X^{([1:N-1])})-r \right)^+ \\
&\quad - (s_{N-1}-s_N) H(X^{([N-2])}|V(\Sc),X^{([N-1:N])}) + \sum_{n=1}^{N-2} (s_n-s_N) H(X^{(n)}|V,X^{([n+1:N])}) \\
&= s_N \left(H(X^{([1:N])})-r\right)^+ + (s_{N-1}-s_N) \left(H(X^{([1:N-1])})-r \right)^+ \\
&\quad + \sum_{n=1}^{N-2} (s_n-s_{N-1}) H(X^{(n)}|V(\Sc),X^{([n+1:N])}),
\end{align*}
where $(a)$ and $(c)$ follow from \eqref{eq:cond_subsetind} and $H(X^{(n)}|V,X^{([n+1:N])})\ge0$ with $n=N$ and $n=N-1$, respectively, and $(b)$ and $(d)$ follow since $(u-v)^+\ge(u)^+-v$ for all $v\ge 0$.
At this point, it is clear that we can apply the same argument for another $N-2$ times and arrive at
\begin{IEEEeqnarray*}{rCl}
R_{\sf u} &\ge& \sum_{n=1}^N (s_n-s_{n+1}) \left(H(X^{([1:n])})-r\right)^+, \\
\label{eq:lower_subsetind}
&=& \sum_{n=1}^N (s_n-s_{n+1}) \left(\sum_{j=1}^nH(X^{(j)})-r\right)^+, \IEEEyesnumber
\end{IEEEeqnarray*}
where $s_{N+1}=0$.
Finally, for independent and identically distributed requests, $s_n(\Sc)=s_n(\ell)$ for all $|\Sc|=\ell$, which concludes the proof of Theorem~\ref{thm:outer-FSN}.

%
%
\section{Proof of Inner Bounds} \label{sec:inner}
In this section, we present the proof of Theorem~\ref{thm:inner-dms}, Proposition~\ref{prop:single-enc-inner}, Theorem~\ref{thm:inner-centralized}, and Theorem~\ref{thm:inner-decentralized}. We begin with the proof of Theorem~\ref{thm:inner-dms}.

The cache contents are formulated by simple digital compressions of the source sequence $x^k$. On the other hand, the update message is formulated by using multiple compressions in which $U^k_\Sc$, $\Sc\subseteq[1:L]$, $\Sc\neq \emptyset$ represents a compression of the pair $(X^k, Y^k)$. The compressions are binned and broadcast through the common link. The destination node $\ell\in[1:L]$ is required to recover only the compressions $U^k_\Sc$ such that $\ell\in\Sc$.

We prove the achievability for $|\Qc|=1$; the rest of the proof follows by time sharing.

\noindent{\bf Rate splitting.} Divide index $m\in[1:2^{nR_{\mathsf{u}}}]$ into $2^L-1$ indices, each indexed by a set $\Sc\subseteq[1:L]$, $\Sc\neq\emptyset$. The indices are denoted by $m_\Sc\in[1:2^{kR_\Sc}]$, $\Sc\subseteq[1:L]$, $\Sc\neq\emptyset$, where $\sum_{\Sc}R_\Sc=R_{\mathsf{u}}$.
\medskip


\noindent{\bf Codebook construction.} Fix a conditional pmf $\prod_{\ell=1}^Lp(v_\ell | x)\prod_{\Sc\subset[1:L]}p(u_\Sc|x, y)$ such that \eqref{eq:comp-cond} is satisfied. To generate a {\em cache codebook} for user $\ell\in[1:L]$, randomly and independently generate $2^{kR_{\mathsf{c}\ell}}$ sequences $v^k_\ell(m_\ell)$, $m_\ell\in[1:2^{kR_{\mathsf{c}\ell}}]$, each according to $\prod_{i=1}^k p(v_{\ell i})$.  To generate the {\em update codebook}, for $\Sc\subseteq[1:L]$, $\Sc\neq\emptyset$, randomly and independently generate $2^{kR_\Sc}$ sequences $u^k_\Sc(m_\Sc, l_\Sc)$, $m_\Sc\in[1:2^{kR_{\Sc}}]$, $l_\Sc\in[1:2^{k\Rh_{\Sc}}]$, each according to $\prod_{i=1}^k p(u_{\Sc i})$.  
Before transmission, the cache codebook for user $\ell$ and the update codebook is revealed to user $\ell\in[1:L]$, and all codebooks are revealed to the encoders.
\medskip

\noindent{\bf Cache encoding.} Upon observing $x^k$, for $\ell\in[1:L]$ the cache encoder finds an index $m_\ell\in[1:2^{kR_{\mathsf{c}\ell}}]$ such that
$(v^k_\ell(m_\ell), x^k)\in\aepvark.$
From the covering lemma~\cite{El-Gamal--Kim2011}, it can be shown that this encoding step is successful with high probability if
\begin{align*}
R_{\mathsf{c}\ell} &> I(V_\ell; X)+\d(\e'), \quad \ell\in[1:L].
\end{align*}
We denote by $M_\ell$, $\ell\in[1:L]$ the index sent to decoder $\ell$ by the cache encoder.
\medskip

\noindent{\bf Update encoding.}
Upon observing $(x^k, y^k)$, for $\Sc\subseteq[1:L]$, $\Sc\neq\emptyset$, the update encoder finds an index pair $(m_\Sc, l_\Sc)\in[1:2^{kR_{\Sc}}]\times [1:2^{k\Rh_{\Sc}}]$ such that
$(u^k_\Sc(m_\Sc, l_\Sc), x^k, y^k)\in\aepvark.$
If there is more than one index pair, select one of them uniformly at random. If there is no such index pair, send an index pair from $[1:2^{kR_{\Sc}}]\times[1:2^{k\Rh_{\Sc}}]$ uniformly at random.
From the covering lemma~\cite{El-Gamal--Kim2011}, it can be shown that this encoding step is successful with high probability if
\begin{align*}
R_{\Sc}+\Rh_{\Sc} > I(U_\Sc; X, Y)+\d(\e'),\quad \Sc\subseteq [1:L], \Sc\neq\emptyset.
\end{align*}
The message $m_\Sc$ is sent to the decoders.
We denote by $M_\Sc$, $\Sc\subseteq[1:L]$, $\Sc\neq\emptyset$ the indices chosen by the update encoder.
\medskip

\noindent{\bf Decoding.}
With $(M_\Sc: \Sc\subseteq[1:L], \Sc\neq \emptyset)$, $y^k$, and $v^k_\ell(M_\ell)$ at hand, decoder $\ell\in[1:L]$ finds the unique index $l_\Sc$ that satisfies
\begin{align*}
(u^k_\Sc(M_\Sc, l_\Sc), y^k, v_\ell^k(M_\ell))\in\aepk,
\end{align*}
for $\Sc$ such that $\ell\in\Sc$.
From the packing lemma~\cite{El-Gamal--Kim2011}, it can be shown that this decoding step is successful with high probability if
\begin{align*}
\Rh_{\Sc} &< I(U_\Sc; Y, V_\ell)-\d(\e), \quad \Sc\subseteq[1:L], \ell\in\Sc.
\end{align*}
By using the fact that $I(U_\Sc; X, Y) = I(U_\Sc; X, Y, V_\ell)$, eliminating the auxiliary rates $R_\Sc$ and $\Rh_\Sc$ with $\sum_\Sc R_\Sc=R_{\mathsf{u}}$ the probability of error for recovering $u^k_\Sc$ tends to zero as $k\to\infty$ if the conditions in Theorem~\ref{thm:inner-dms} are satisfied. Finally, since we choose a joint distribution that satisfies condition~\eqref{eq:ergodic-cond} and by the typical average lemma~\cite{El-Gamal--Kim2011}, the probability of error tends to zero as $k\to\infty$.

\begin{remark}
The decoding phase for the update messages can be further improved by applying some decoding order on $U^k_\Sc$ such that receiver $\ell\in\Sc$, $\ell\in\Sc'$ decodes $U^k_\Sc$ before $U^k_{\Sc'}$ for every $|\Sc|>|\Sc'|$.
By this ordering, when decoding $U^k_{\Sc'}$, the decoder can further use $U_\Sc$ as {\em side information} which results in the condition
\begin{align*}
R_\mathsf{u} &> \sum_{\substack{\Sc\subseteq[1:L]}}\max_{\ell\in\Sc} I(U_{\Sc}; X| (U_{\Sc'}: \ell\in\Sc', |\Sc'|>|\Sc|),V_\ell, Y, Q).
\end{align*}
\end{remark}

Next, to prove Proposition~\ref{prop:single-enc-inner} which applies to the static requests and the single block encoding case, we only need some minor modifications from the above steps in which we highlight in the following. For the cache encoder, we follow the same encoding step as in the previous case since for both cases, the cache encoder does not depend on the request information. As for the update stage, we fix a distribution $\prod_{\Sc}p(u_\Sc|x, y)$. For $\Sc\subseteq[1:L]$, $\Sc\neq\emptyset$, randomly and independently generate $2^{kR_{\Sc, y}}$ sequences $u^k_{\Sc}(m_{\Sc, y}, l_{\Sc, y})$, $m_{\Sc,y}\in[1:2^{kR_{\Sc,y}}]$, $l_{\Sc,y}\in[1:2^{k\Rh_{\Sc,y}}]$, each according to $\prod_{i=1}^k p(u_{\Sc i}|y)$, where $\sum_{\Sc}R_{\Sc,y}=R_{\sf u}(y)$. Upon observing $(x^k, y)$, for $\Sc\subseteq[1:L]$, $\Sc\neq\emptyset$, the update encoder finds an index pair $(m_{\Sc,y}, l_{\Sc,y})$ such that
$(u^k_\Sc(m_{\Sc,y}, l_{\Sc,y}), x^k)\in\aepvark(U_\Sc, X)$,
where the typical set $\aepvark(U_\Sc, X)$ is defined over $p(u_\Sc, x|y)$.
This step is successful with high probability if
\begin{align*}
R_{\Sc,y}+\Rh_{\Sc,y} > I(U_\Sc; X| Y=y)+\d(\e').
\end{align*}
At the decoder, with $M_{\Sc,y}$, $y$, and $v^k_\ell(M_\ell)$ at hand, decoder $\ell\in[1:L]$ finds the unique index $l_{\Sc,y}$ that satisfies
\begin{align*}
(u^k_\Sc(M_{\Sc,y}, l_{\Sc,y}), v_\ell^k(M_\ell))\in\aepk(U_\Sc, V),
\end{align*}
for $\Sc$ such that $\ell\in\Sc$, where the typical set $\aepk(U_\Sc, V)$ is defined over $p(u_\Sc, v|y)$.
This decoding step is successful with high probability if
\begin{align*}
\Rh_{\Sc} &< I(U_\Sc; V_\ell|Y=y)-\d(\e), \quad \Sc\subseteq[1:L], \ell\in\Sc.
\end{align*}
By eliminating the auxiliary rates $R_{\Sc,y}$ and $\Rh_{\Sc,y}$ with $\sum_\Sc R_{\Sc,y}=R_{\mathsf{u}}(y)$, we arrive at the conditions in Proposition~\ref{prop:single-enc-inner}.

In the next subsections, we specify the choice of auxiliary random variables to characterize achievable rate regions for FSNs. The use of coded time sharing is critical in the analysis.

%
%
\subsection{Proof of Theorem~\ref{thm:inner-centralized} and Corollary~\ref{cor:centralized-unif}} \label{sec:centralized}
We show the rate--cache tradeoff for $R_{\mathsf{c}}=0, \frac{1}{L},\frac{2}{L},\ldots, N$. Fix $r_n\in[0:L]$, $n\in[1:N]$ such that $\sum_{n=1}^Nr_n=LR_{\mathsf{c}}$.
The auxiliary random variables in Theorem~\ref{thm:inner-dms} are chosen as follows.
Let $Q=(Q_n: n\in[1:N])$, where $\Qc_n=\{\Tc_n: \Tc_n\subset[1:L], |\Tc_n|=r_n\}$ and $Q_n\sim \U\left(\Qc_n\right)$.
For $n\in[1:N]$, $\Tc_n\subseteq[1:L]$, $|\Tc_n|=r_n$, define
\begin{align}
W^{(n)}_{\Tc_n}=X^{(n)}\cdot \mathbb{1}\{Q_n=\Tc_n\}, \label{eq:def-w}
\end{align}
where $\mathbb{1}\{\Ac\}$ is the indicator function of the event $\Ac$.
The auxiliary random variables $V_\ell$, $\ell\in[1:L]$ and $U_\Sc$, $\Sc\subseteq[1:L]$, $\Sc\neq\emptyset$ are chosen as a collection of $W^{(n)}_{\Tc_n}$. For $\ell\in[1:L]$, we choose
\begin{align}
V_\ell=\left(W^{(n)}_{\Tc_n}: n\in[1:N], {\Tc_n}\subseteq[1:L], |{\Tc_n}|=r_n, \ell\in{\Tc_n}\right). \label{eq:def-vell}
\end{align}
On the other hand, for $\Sc\subseteq[1:L]$, $\Sc\neq\emptyset$, we choose
\begin{align}
U_\Sc=\left(W^{(n)}_{\Tc_n}\cdot A_\Sc(\Tc_n, n): n\in[1:N], {\Tc_n}\subseteq[1:L], |{\Tc_n}|=r_n\right) \label{eq:def-ues}
\end{align}
where
\begin{align}
\Ac_{\Sc}(\Tc_n, n) = \mathbb{1}\left\{\Tc_n\subset \Sc\right\}\left(\prod_{j\in\Sc\setminus\Tc_n}\mathbb{1}\left\{Y_j=n\right\}\right)\left(\prod_{j\in\Sc^c}\mathbb{1}\left\{Y_j\neq n \right\}\right), \label{eq:Aes}
\end{align}
and $\Sc^c=[1:L]\setminus \Sc$.
Note that the above choice of auxiliary random variables satisfy~\eqref{eq:ergodic-cond}.

With the above choice, the cache rate is given by
\begin{align}
R_{\mathsf{c}\ell}&> I(V_\ell; X|Q)\nn\\
&= H(V_\ell | Q) \nn\\
&= \sum_{n=1}^N \sum_{\substack{{\Tc_n}\subseteq[1:L]: \ell\in{\Tc_n},\\ |{\Tc_n}|=r_n}} H(W_{\Tc_n}^{(n)} | Q_n) \nn\\
&= \sum_{n=1}^N \sum_{\substack{{\Tc_n}\subseteq[1:L]: \ell\in{\Tc_n},\\ |{\Tc_n}|=r_n}} \frac{1}{\binom{L}{r_n}} \nn\\
&= \sum_{n=1}^N \binom{L-1}{r_n-1}\frac{1}{\binom{L}{r_n}} \nn\\
&= \sum_{n=1}^N \frac{r_n}{L}= R_{\mathsf{c}}. \label{eq:cache-rate}
\end{align}

On the other hand, note that $I(U_\Sc; X|V_\ell, Y, Q)=H(U_\Sc|V_\ell, Y, Q)$, and
\begin{align*}
&H(U_\Sc|V_{\ell}, Y, Q) \\
&= \sum_{n=1}^N \sum_{\Tc_n: |\Tc_n|=r_n} H(W^{(n)}_{\Tc_n} \cdot A_\Sc(\Tc_n, n)|V_\ell, Y, Q)\\
&\stackrel{(a)}{=} \sum_{n=1}^N \sum_{\Tc_n: |\Tc_n|=r_n,\ell\not\in\Tc_n} H(W^{(n)}_{\Tc_n} \cdot A_\Sc(\Tc_n, n)|Y, Q)\\
&= \sum_{n=1}^N \sum_{\Tc_n: |\Tc_n|=r_n,\ell\not\in\Tc_n} \frac{1}{\binom{L}{r_n}} H(W^{(n)}_{\Tc_n} \cdot A_\Sc(\Tc_n, n)|Y, Q, Q_n=\Tc_n)\\
&= \sum_{n=1}^N \sum_{\Tc_n: |\Tc_n|=r_n,\ell\not\in\Tc_n} \frac{1}{\binom{L}{r_n}} \sum_{y\in\Yc}p_Y(y) H(W^{(n)}_{\Tc_n} \cdot A_\Sc(\Tc_n, n)|Y=y, Q, Q_n=\Tc_n)\\
&= \sum_{n=1}^N \sum_{\Tc_n: |\Tc_n|=r_n,\ell\not\in\Tc_n} \frac{1}{\binom{L}{r_n}} \sum_{y\in\Yc}p_Y(y) \mathbb{1}\left\{\Tc_n\subset \Sc\right\}\left(\prod_{j\in\Sc\setminus\Tc_n}\mathbb{1}\left\{y_j=n\right\}\right)\left(\prod_{j\in\Sc^c}\mathbb{1}\left\{y_j\neq n \right\}\right)\\
&\stackrel{(b)}{=} \sum_{n=1}^N \sum_{\substack{\Tc_n: |\Tc_n|=r_n,\\ \ell\not\in\Tc_n, \Tc_n\subset\Sc}} \frac{1}{\binom{L}{r_n}} p_n^{|\Sc|-r_n}(1-p_n)^{|\Sc^c|},
\end{align*}
where $p_n=\P\{Y_1=n\}$, step $(a)$ follows since $W_{\Tc_n}^{(n)}\in V_\ell$ for $\ell \in \Tc_n$, and step $(b)$ follows since $p_Y(y)=\prod_{j=1}^L p_{Y_1}(y_j)$, $p_{Y_1}(n)=p_n$.
Let $\psi_n(\Sc)=p_n^{|\Sc|-r_n}(1-p_n)^{|\Sc^c|}$. Then, the update rate can be evaluated by
\begin{align*}
R_{\mathsf{u}}&> \sum_{\Sc\subseteq[1:L]}\max_{\ell\in\Sc}H(U_\Sc|V_{\ell}, Y, Q)\\
&= \sum_{\Sc\subseteq[1:L]}\max_{\ell\in\Sc}\sum_{n=1}^N \sum_{\substack{\Tc_n: |\Tc_n|=r_n,\\ \ell\not\in\Tc_n, \Tc_n\subset\Sc}} \frac{1}{\binom{L}{r_n}} \psi_n(\Sc)\\
&= \sum_{\Sc\subseteq[1:L]}\max_{\ell\in\Sc}\sum_{n=1}^N \mathbb{1}\{|\Sc|>r_n\}  \frac{\binom{|\Sc|-1}{r_n}}{\binom{L}{r_n}} \psi_n(\Sc)\\
&= \sum_{\Sc\subseteq[1:L]}\sum_{n=1}^N \mathbb{1}\{|\Sc|>r_n\}  \frac{\binom{|\Sc|-1}{r_n}}{\binom{L}{r_n}} \psi_n(\Sc)\\
&= \sum_{j=1}^L\sum_{\Sc:|\Sc|=j}\sum_{n=1}^N \mathbb{1}\{|\Sc|>r_n\}  \frac{\binom{|\Sc|-1}{r_n}}{\binom{L}{r_n}} \psi_n(\Sc)\\
&= \sum_{j=1}^L\sum_{\Sc:|\Sc|=j}\sum_{n=1}^N \mathbb{1}\{j>r_n\}  \frac{\binom{j-1}{r_n}}{\binom{L}{r_n}} p_n^{j-r_n}(1-p_n)^{L-j}\\
&= \sum_{n=1}^N\sum_{j=1}^L \mathbb{1}\{j>r_n\} \binom{L}{j} \frac{\binom{j-1}{r_n}}{\binom{L}{r_n}} p_n^{j-r_n}(1-p_n)^{L-j}\\
&= \sum_{n=1}^N\sum_{j=r_n+1}^L  \binom{L}{j} \frac{\binom{j-1}{r_n}}{\binom{L}{r_n}} p_n^{j-r_n}(1-p_n)^{L-j}\\
&=\sum_{n=1}^N\sum_{j=1}^{L-r_n}\frac{j}{j+r_n}  \binom{L-r_n}{j} p_n^{j}(1-p_n)^{L-r_n-j}.
\end{align*}
This concludes the proof of Theorem~\ref{thm:inner-centralized}.

Next, specializing to uniform requests, let $r_n=r$ such that $r=L\Rcache/N$. Then,
\begin{align*}
R_{\mathsf{u}}&> \sum_{n=1}^N\sum_{j=1}^{L-r_n}\frac{j}{j+r_n}  \binom{L-r_n}{j} p_n^{j}(1-p_n)^{L-r_n-j}\\
&=N\sum_{j=1}^{L-r}\frac{j}{j+r}  \binom{L-r}{j} \left(\frac{1}{N}\right)^{j}\left(1-\frac{1}{N}\right)^{L-r-j}.
\end{align*}
Thus, for $\Rcache=0$, we have $r=0$, which gives
\begin{align*}
R_{\mathsf{u}}&> N\sum_{j=1}^{L} \binom{L}{j} \left(\frac{1}{N}\right)^{j}\left(1-\frac{1}{N}\right)^{L-j}\\
&=N\sum_{j=0}^{L} \binom{L}{j} \left(\frac{1}{N}\right)^{j}\left(1-\frac{1}{N}\right)^{L-j} - N\left(1-\frac{1}{N}\right)^{L}\\
&=N(1-(1-1/N)^L).
\end{align*}
For $r\in[1:L]$ such that $r=LR_{\mathsf{c}}/N$, we have
\begin{align*}
R_{\mathsf{u}}&> N\sum_{j=1}^{L-r}\frac{j}{j+r}  \binom{L-r}{j} \left(\frac{1}{N}\right)^{j}\left(1-\frac{1}{N}\right)^{L-r-j}\\
&= N\sum_{j=0}^{L-r}\frac{j}{j+r}  \binom{L-r}{j} \left(\frac{1}{N}\right)^{j}\left(1-\frac{1}{N}\right)^{L-r-j}\\
&= \E\left[\frac{Z}{Z+r}\right],
\end{align*}
where $Z\sim\text{Binom}(L-r,1/N)$.

%
%
\subsection{Proof of Theorem~\ref{thm:inner-decentralized}}\label{sec:decentralized}
Consider any cache rate $R_{\mathsf{c}}\in[0,N]$ and let $r_n\in[0,1]$ such that $\sum_{n=1}^N r_n =R_{\mathsf{c}}$. The auxiliary random variables are chosen in the following manner.
Let $Q=\{Q^{(n)}_\ell: n\in[1:N], \ell\in[1:L]\}$, where $Q^{(n)}_\ell$ are independent of each other and $Q_\ell^{(n)}\sim\Bern(r_n)$.
For $\ell\in[1:L]$, we choose
\begin{align}
V_\ell = \left(X^{(n)}Q_\ell^{(n)}:n\in[1:N]\right). \label{eq:vell-decentralized}
\end{align}
Note that with this particular choice of $V_\ell$, the caching strategy is {\em decentralized}.
On the other hand, for $\mathcal{S}\subseteq [1:L]$, $S\neq \emptyset$, we choose
\begin{align*}
U_\mathcal{S} = \left(X^{(n)}\cdot A_{\mathcal{S},n}:n\in[1:N]\right), \label{eq:ues-decentralized}
\end{align*}
where
\begin{align*}
A_{\mathcal{S},n} &= \prod_{j\in S}\mathbb{1}\{Y_j=n \text { or } Q_j^{(n)} = 1\}\prod_{j\in S^c}\mathbb{1}\{Y_j\neq n \text { and } Q_j^{(n)} = 0\}.
\end{align*}
Note that the above choice of auxiliary random variables satisfy~\eqref{eq:ergodic-cond}.
Then, the cache rate is given by
\begin{IEEEeqnarray*}{rCl}
R_{\mathsf{c}\ell} &>& I(X;V_\ell|Y,Q)\\
&=& H(V_\ell|Q) \\
&=& \sum_{n=1}^N \P(Q_\ell^{(n)}=1) \\
&=&  \sum_{n=1}^N r_n = R_{\sf c}.
\end{IEEEeqnarray*}
Furthermore, for $\mathcal{S}\subseteq[1:L]$ and $\ell\in\mathcal{S}$, we have
\begin{align*}
&H(X^{(n)}\cdot A_{\mathcal{S},n}|V_\ell,Y,Q) \\
&= H(X^{(n)}\cdot A_{\mathcal{S},n}|(X^{(n)}\cdot Q_\ell^{(n)}),Y,Q)\\
&= \P\{Q_\ell^{(n)}=0\} H(A_{\mathcal{S},n}X^{(n)}|(X^{(n)}\cdot Q_\ell^{(n)}),Y,Q,Q_\ell^{(n)}=0) \\
&= \P\{Q_\ell^{(n)}=0\} \P\{Y_\ell=n\} \prod_{j\in\mathcal{S}\backslash\{\ell\}}
(1-p_{j n}) \prod_{j\in\mathcal{S}^c} p_{j n} \\
&= \left(1-r_n\right)p_n(1-\alpha_n)^{|\mathcal{S}|-1}\alpha_n^{|\mathcal{S}^c|},
\end{align*}
where $p_{j n} = \P\{Y_j\neq n \text { and } Q_j^{(n)} = 0\}$ and $\alpha_n = (1-p_n)(1-r_n)$.
Thus, the update rate is given by
\begin{align}
R_{\sf u} &> \sum_{\mathcal{S}\subseteq[1:L]} \max_{\ell\in\mathcal{S}} H(U_{\mathcal{S}}|V_\ell,Y,Q) \nn\\
&= \sum_{j=1}^L \sum_{\mathcal{S}:|\mathcal{S}|=j} \sum_{n=1}^N \left(1-r_n\right)p_n(1-\alpha_n)^{|\mathcal{S}|-1}\alpha_n^{|\mathcal{S}^c|} \nn\\
&= \sum_{j=1}^L \binom{L}{j} \sum_{n=1}^N \left(1-r_n\right)p_n(1-\alpha_n)^{j-1}\alpha_n^{L-j}  \nn\\
&= \sum_{n=1}^N \frac{p_n\left(1-r_n\right)}{1-\alpha_n} \sum_{j=1}^L \binom{L}{j} (1-\alpha_n)^{j}\alpha_n^{L-j} \\
&= \sum_{n=1}^N \frac{p_n\left(1-r_n\right)}{1-\alpha_n} (1-\alpha_n^L). \label{eq:geo}
\end{align}
This concludes the proof of Theorem~\ref{thm:inner-decentralized}.

%
%
\section{Numerical Evaluations} \label{sec:numerical}
In this section, we provide an algorithm for numerically optimizing Theorem~\ref{thm:inner-centralized}, some notes on the  optimization of Theorem~\ref{thm:inner-decentralized}, and some numerical examples of the outer bound and the centralized and the decentralized inner bounds.

We begin by providing an optimization algorithm for Theorem~\ref{thm:inner-centralized}.
\begin{proposition}\label{prop:greedy}
For $R_{\mathsf{c}}=0,\frac{1}{L},\frac{2}{L},\ldots,N$, 
Algorithm \ref{alg:greedy} finds the minimum value of $R_{\sf u}(\Rcache)$ for the centralized strategy in Theorem~\ref{thm:inner-centralized}, where $Z_n\sim \text{Binom}(L-r_n,p_n)$.
\end{proposition}
The proof of this proposition is given in Appendix~\ref{app:greedy}.
\begin{algorithm}
\caption{Greedy Algorithm}\label{alg:greedy}
\begin{algorithmic}
\State Initialization:
\State $\Nc \gets [1:N]$;
\State $\Rcache \gets 0$;
\For {$n = 1,\cdots, N$}
\State $r_n \gets 0$;
\State $e_n \gets p_n\left(1-(1-p_n)^L\right) - \E_{Z_n}\left[\frac{Z_n}{Z_n+1}\right]$;
\EndFor
\State $R \gets \sum_{n=1}^N p_n\left(1-(1-p_n)^L\right)$;
\State $R_{\sf u}(\Rcache) \gets R$;

\For{$\Rcache = \frac{1}{L}, \frac{2}{L}, \cdots, N-\frac{1}{L}, N$}
\State $m \gets \displaystyle\argmax_{n\in\Nc} e_n$;
\State $R \gets R - e_m$;
\State $r_m \gets r_m + 1$;
\If {$r_m = L$}
\State $\Nc \gets \Nc\backslash\{m\}$;
\Else
\State $e_m = \E_{Z_n}\left[\frac{Z_n}{Z_n+r_m+1}\right] - \E_{Z_n}\left[\frac{Z_n}{Z_n+r_m}\right]$;
\EndIf
\State $R_{\sf u}(\Rcache)=R$;
\EndFor\\
\Return $R_{\sf u}$
\end{algorithmic}
\end{algorithm}

Next, we consider the decentralized strategy in Theorem~\ref{thm:inner-decentralized}. Then, for $R_{\sf c}\in [0,N]$, finding the minimum rate--cache tradeoff for the right hand side of equation \eqref{eq:inner-decentralized} requires optimization over $r_n\in[0,1]$ such that $\sum_{n=1}^N r_n=\Rcache$. The process can be cast as the following {\em convex optimization problem}~\cite{Boyd--Vandenberghe2004}:
\begin{align*}
\text{minimize} 
&\qquad \sum_{n=1}^N p_n \sum_{\ell=0}^{L-1} \left(1-p_n\right)^\ell\left(1-r_n\right)^{\ell+1}, \\
\text{subject to} &\qquad 0 \le r_n \le 1, \quad \forall n\in[1:N], \\
&\qquad  \sum_{n=1}^N r_n = R_{\sf c}.
\end{align*}
For the following discussion, we assume that $p_n\in(0,1)$ for all $n\in[1:N]$ and $L\ge 2$. Now let us consider the Lagrange function
\begin{align*}
& \mathcal{L}(\rv, \mu, \nu,\lambda) \\
&=  \sum_{n=1}^N p_n \sum_{\ell=0}^{L-1} \left(1-p_n\right)^\ell\left(1-r_n\right)^{\ell+1} + \sum_{n=1}^N \mu_n(-r_n)  + \sum_{n=1}^N \nu_n(r_n-1) + \lambda\left(\sum_{n=1}^Nr_n-R_{\sf c}\right),
\end{align*}
where $\rv=(r_1,\ldots, r_N)$, $\mu=(\mu_1,\ldots, \mu_N)$, and $\nu=(\nu_1,\ldots, \nu_N)$.
Denote by $\rv^\star$ and $(\mu^\star,\nu^\star,\lambda^\star)$ the optimal solutions for the primal and dual problems, respectively.
Since the optimization problem is convex, the corresponding Karush--Kuh--Tucker (KKT) conditions are sufficient for optimality. In particular, we have for $n\in[1:N]$,
\begin{enumerate}
\item $r_n = 1$ if and only if $p_n \ge \lambda^\star$;
\item $r_n = 0$ if and only if
\begin{align*}
p_n\sum_{\ell=0}^{L-1}(\ell+1)(1-p_n)^\ell &\le \lambda^\star;
\end{align*}
\item $r_n \in (0,1)$ if and only if
\begin{align*}
p_n\sum_{\ell=0}^{L-1}(\ell+1)(1-p_n)^\ell(1-r_n)^\ell &= \lambda^\star.
\end{align*}
\end{enumerate}
%

In the following, we compare the centralized and decentralized inner bounds with an  {\em uncoded} baseline strategy which follows the principle of caching the highest popularity first (HPF).
In~\cite{Wang--Lim--Gastpar2015}, it was shown that HPF is optimal for the single user FSN. The HPF achievable rate pair for the multi-user network is given by
\begin{align}
R_{\mathsf{HPF}}(R_{\mathsf{c}}) &> \sum_{n=R_{\mathsf{c}}+1}^N (1-(1-p_n)^{L})\label{eq:hpf}
\end{align}
for $R_{\mathsf{c}}\in[0:N]$.

For numerical examples, we consider a Zipf distribution on the file popularities, i.e., the popularity of file $n\in[1:N]$ is given by
\begin{align*}
p_n=\frac{n^{-\alpha}}{\sum_{\tilde{n}=1}^N \tilde{n}^{-\alpha}},
\end{align*}
for some fixed parameter $\alpha\ge 0$.

\begin{figure}[t!]
\begin{center}
\footnotesize
\psfrag{a}[c]{$R_{\mathsf{c}}$}
\psfrag{b}[c]{$R_{\mathsf{u}}$}
\hspace{6pt}\includegraphics[width=0.55\textwidth]{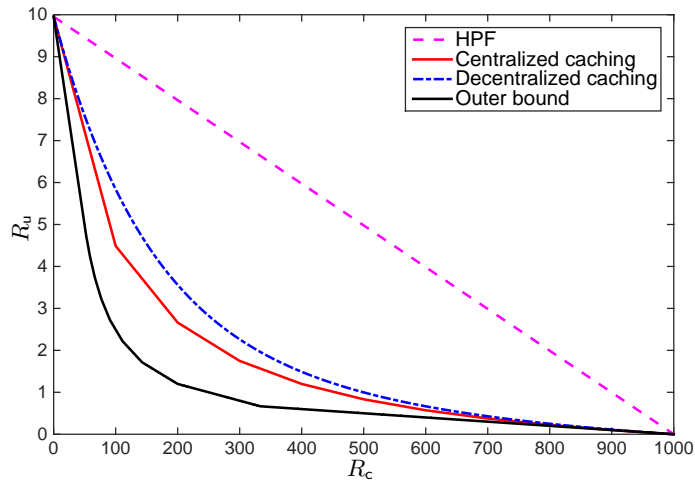}
\caption{The cache--rate tradeoff curves for the centralized scheme (upper solid curve), the decentralized scheme (dash-dotted curve), the outer bound (bottom solid curve), and the HPF strategy (dashed curve) for $N=1000$, $L=10$, and $\alpha=0$.}
\label{fig:unif1}
\end{center}
\vspace{-1em}
\end{figure}

\begin{figure}[t!]
\begin{center}
\footnotesize
\psfrag{a}[c]{$R_{\mathsf{c}}$}
\psfrag{b}[b]{$R_{\mathsf{u}}$}
\hspace{6pt}\includegraphics[width=0.55\textwidth]{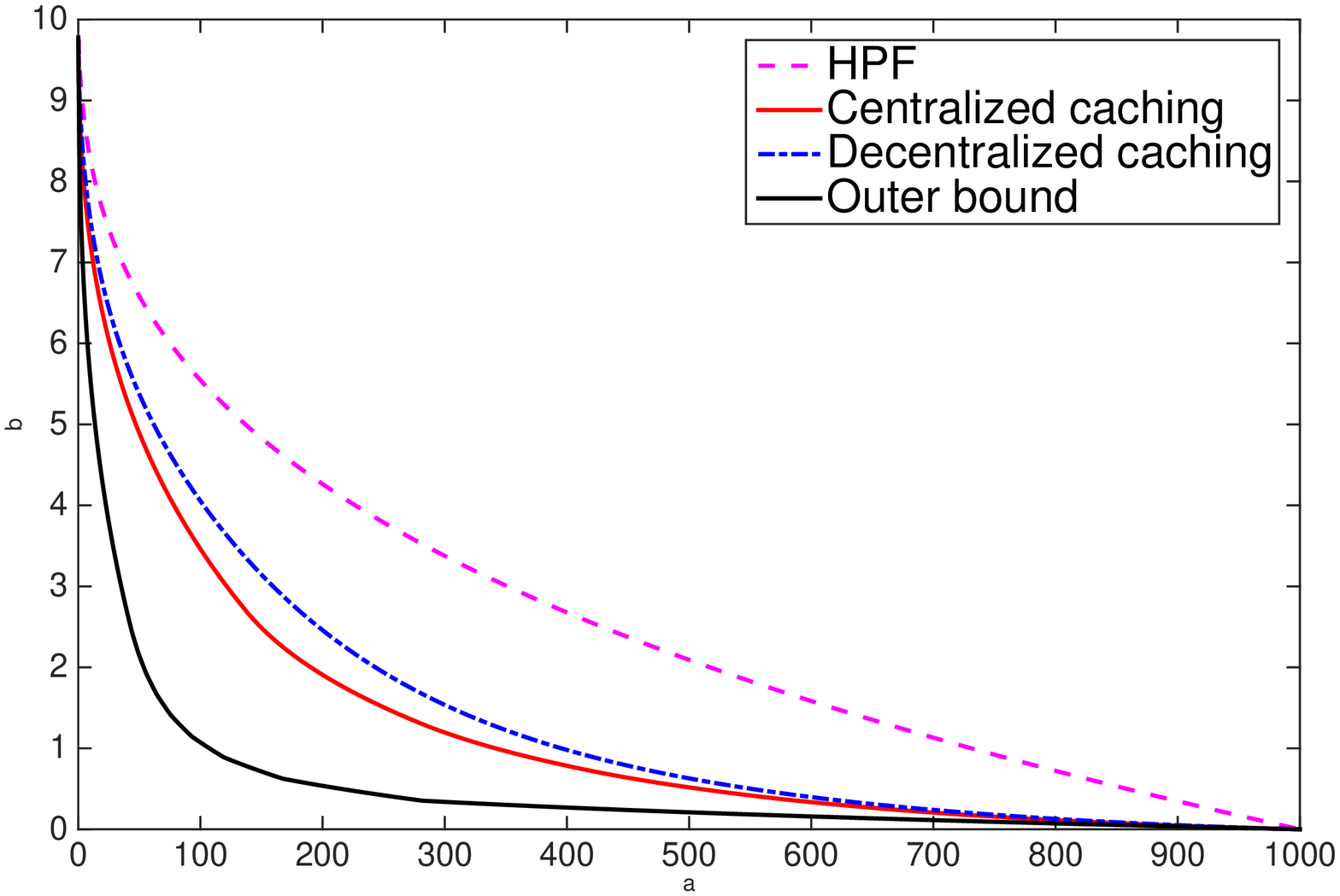}
\caption{The cache--rate tradeoff curves for the centralized scheme (upper solid curve), the decentralized scheme (dash-dotted curve), the outer bound (bottom solid curve), and the HPF strategy (dashed curve) for $N=1000$, $L=10$, and $\alpha=0.7$.}
\label{fig:alpha1}
\end{center}
\vspace{-1em}
\end{figure}

\begin{figure}[t!]
\begin{center}
\footnotesize
\psfrag{a}[c]{$R_{\mathsf{c}}$}
\psfrag{b}[c]{$R_{\mathsf{u}}$}
\hspace{6pt}\includegraphics[width=0.55\textwidth]{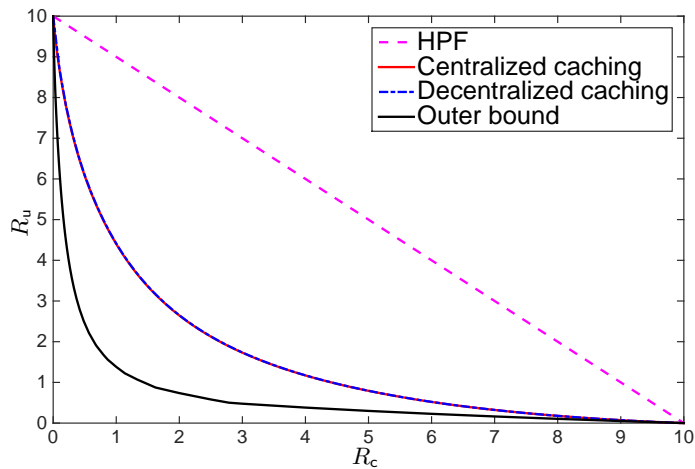}
\caption{The cache--rate tradeoff for the centralized and decentralized schemes (closely merged in the upper solid curve), the outer bound (bottom solid curve), and the HPF strategy (dashed curve) for $N=10$, $L=1000$, and $\alpha=0.7$.}
\label{fig:alpha2}
\end{center}
\vspace{-1em}
\end{figure}

In Figure~\ref{fig:unif1} we compare the performance of the two inner bounds, the HPF strategy, and the outer bound for the case $N=1000$, $L=10$, and $\alpha=0$, i.e., the case when the files are uniformly distributed.
In Figures~\ref{fig:alpha1} and~\ref{fig:alpha2},  we compare the inner bounds and the outer bound for the cases $\alpha=1.2$ with $(N=1000, L=10)$ and $(N=10, L=1000)$, respectively.
In all cases, the inner bounds in Theorems~\ref{thm:inner-centralized} and~\ref{thm:inner-decentralized} are within a constant multiplicative factor of $4$ from the outer bound in Theorem~\ref{thm:outer-FSN}. On the other hand, the HPF strategy shows poor performance when the users' requests become uniformly distributed or the number of users is large compared to the number of files.

%
%
\section{Concluding Remarks} \label{discussion}
Following up on our previous information theoretic approach that formulated single and two-user cache aided networks in terms of a distributed source coding problem, in this paper, we have extended the approach and provided inner and outer bounds for several cache networks with multiple users.

Looking back, there has been several diverse approaches that have been taken to understand the benefit of coded caching, e.g., distributed source coding~\cite{Wang--Lim--Gastpar2015, Timo--Bidokhti--Wigger--Geiger2016}, network coding~\cite{Maddah-Ali--Niesen2014, Maddah-Ali--Niesen2015}, computational~\cite{Tian2015}, and index coding~\cite{Ji14m, Ji14z, Wan--Tuninetti--Piantanida2015} based approaches have been developed. Compared to the distributed source coding approach which is based on random coding arguments, the advantage of a (linear) network coding approach is that it explicitly reveals the coding strategy with potentially lower complexity. On the other hand, in terms of theoretical analysis on the performance of these coding techniques,
as originally developed in the paper by Ahlswede, Cai, Li, and Yeung~\cite{Ahlswede--Cai--Li--Yeung2000},
network coding can be specialized from the more general random coding theorems, e.g., \cite{Lim--Kim--El-Gamal--Chung2011}. We have also demonstrated this by showing that our coding theorem based on random coding arguments can recover the network coding based strategies which is accomplished by substituting network coding with random binning.

On the other hand, the relation with index coding based approaches is less apparent. The idea of translating the cache network into an index coding problem is as follows. Under the assumption that the cache content is fixed to some fraction of the database (in a uncoded fashion), and assuming that the users' requests are fixed, the update phase can be viewed as an index coding problem. In general, the uncoded cache placement assumption itself may lead to a sub-optimal strategy for the caching problem. Nonetheless, several approaches adopt this assumption, including our choice of auxiliary random variables which enables the analysis to be more tractable and in several cases is sufficient to obtain order optimality.
Under such assumptions, there is an interesting analogy with the index coding results in~\cite{Arbabjolfaei--Bandemer--Kim--Sasoglu--Wang2013}. In~\cite{Arbabjolfaei--Bandemer--Kim--Sasoglu--Wang2013}, the authors provide an achievable scheme based on random coding for the index coding problem instead of the more commonly used graph theoretic, algebraic, and network coding based approaches. Using this approach, the authors showed that a composite random coding strategy is optimal for all index coding problems with up to five messages. Our update coding strategy is reminiscent of this composite coding strategy in that it is represented by the auxiliary random variables $U_\Sc$, $\Sc\subseteq[1:L]$, $\Sc\neq\emptyset$, for which only the decoders in $\ell\in\Sc$ recovers $U_\Sc$. However, in general, the composite coding strategy can be strictly suboptimal for index coding. It would be interesting further work to seek for improved strategies over our proposed composite coding strategy for cache aided networks.

%
%
\appendices
%
%
\section{Analysis of Multiplicative Gap Results}
\subsection{Proof of Theorem~\ref{thm:gap-ergodic}} \label{app:gap-erogdic}
Denote the right hand side of \eqref{eq:dec-unif} by $\bar{R}_{\sf u\tdash dc}(R_{\mathsf{c}})$.
Note that we have $\bar{R}_{\sf u\tdash dc}(0)=R_{\sf u}^\star(0)$.
To prove Corollary~\ref{cor:decentralized-unif}, we consider the following (relaxed) achievable rate--cache region given by the convex hull of the point $(R_{\sf c},R_{\sf u})=(0,R^*_{\sf u}(0))$ and the set
\begin{align*}
\left\{(R_{\sf c},R_{\sf u}) : R_{\sf u} \ge \frac{N-R_{\sf c}}{1+R_{\sf c}(1-1/N)}, R_{\sf c} \in [0,N]\right\}.
\end{align*}
Denote by $\breve{R}_{\sf u\tdash dc}(R_{\sf c})$ the corresponding rate region.
Now, we show that given a fixed cache rate $R_{\sf c}\ge0$, the decentralized coded caching scheme in Corollary \ref{cor:decentralized-unif} achieves an update rate within a constant multiplicative factor from the rate--cache function $R_{\sf u}^\star(R_{\sf c})$ for uniform requests. Let $R_{\sf u\tdash lb}(R_{\sf c})$ denote the right hand side of~\eqref{eq:outer_uni}.

Since $R_{\sf u\tdash lb}(R_{\sf c}) \le R_{\sf u}^\star(R_{\sf c})$, it suffices to show that $\frac{\breve{R}_{\sf u\tdash dc}(R_{\sf c})}{R_{\sf u \tdash lb}(R_{\sf c})} \le 4$ for $\Rcache\in[0,N)$.
If $N=1$, it can be easily checked that $\breve{R}_{\sf u\tdash dc}(R_{\sf c})=R_{\sf u \tdash lb}(R_{\sf c})=1-\Rcache$. In the following, we assume that $N\ge 2$.
For notational convenience, we denote $\overline{L} = \min\{L,N\}$. The lower bound $R_{\sf u \tdash lb}(R_{\sf c})$ is an intersection of half planes, and the corner points of $R_{\sf u \tdash lb}(R_{\sf c})$ are characterized by the set $\Omega = \{(\omega_\ell,R_{\sf u \tdash lb}(R_{\sf c})):\ell\in\{0,1,\cdots,\overline{L}\}\}$, where
\begin{IEEEeqnarray*}{rCl}
\omega_\ell &:=& \begin{cases}
N & \text{ if } \ell=0, \\
\frac{N\left(1-\frac{1}{N}\right)^\ell}{N+(\ell+1-N)\left(1-\frac{1}{N}\right)^\ell} & \text{ if } \ell\in[\overline{L}-1], \\
0 & \text{ if } \ell=\overline{L}.
\end{cases}
\end{IEEEeqnarray*}
We note that for $\ell\in[1:\overline{L}-1]$, the two lines $y=(1-(1-1/N)^\ell)(N-\ell x)$ and $y=(1-(1-1/N)^{\ell+1})(N-(\ell+1) x)$ intersect at $x=\omega_\ell$.

Next, we relax the inner bound $\breve{R}_{\sf u\tdash dc}(R_{\sf c})$ by the following piecewise-linear bound resulting from $\Omega$:
\begin{align*}
\breve{R}'_{\sf u\tdash dc}(R_{\sf c}) &:= (1-\theta)\breve{R}_{\sf u\tdash dc}(\omega_{\ell}) + \breve{R}_{\sf u\tdash dc}(\omega_{\ell-1}),
\end{align*}
if $\Rcache = (1-\theta)\omega_{\ell} + \theta\omega_{\ell-1}$ for some $\theta\in [0,1)$, $\ell\in[1:\overline{L}]$. Note that $\breve{R}'_{\sf u\tdash dc}(R_{\sf c})
= \breve{R}_{\sf u\tdash dc}(R_{\sf c})$ for all $\Rcache\in\{\omega_0,\omega_1,\cdots,\omega_{\overline{L}}\}$. Then, for each segment $[\omega_{\ell},\omega_{\ell-1})$, $\ell\in[1:\overline{L}]$, the ratio $\frac{\breve{R}'_{\sf u\tdash dc}(R_{\sf c}) }{R_{\sf u\tdash lb}(\Rcache)}$ is a linear-fractional function with respect to $\Rcache$ and thus is quasiconvex \cite{Boyd--Vandenberghe2004}. A quasiconvex function has the property that the value of the function on a segment does not exceed the maximum of its values at the endpoints. Therefore, it suffices to check whether $\frac{\breve{R}_{\sf u\tdash dc}(R_{\sf c}) }{R_{\sf u\tdash lb}(\Rcache)}\le 4$ for all $\Rcache\in\{\omega_0,\omega_1,\cdots,\omega_{\overline{L}}\}$. First, it is clear that we have
\begin{align*}
\lim_{\Rcache\to N^-}\frac{\breve{R}_{\sf u\tdash dc}(R_{\sf c})}{R_{\sf u\tdash lb}(\Rcache)} &= 1.
\end{align*}
Also, we have
\begin{align*}
\frac{\breve{R}_{\sf u\tdash dc}(R_{\sf c})}{R_{\sf u\tdash lb}(\Rcache)}  &= \frac{N(1-(1-1/N)^L)}{(1-(1-1/N)^{\overline{L}})N} \\
&= \frac{1-(1-1/N)^L}{1-(1-1/N)^{\min\{L,N\}}} \\
&= \begin{cases}
\frac{1-(1-1/N)^L}{1-(1-1/N)^N} & \text{ if }  L > N \\
1 & \text{ if }  L\le N
\end{cases} \\
&\stackrel{(a)}{\le} \frac{1}{1-e^{-1}} \approx 1.5820,
\end{align*}
where $(a)$ follows since $(1-1/z)^z \le e^{-1}$ for all $z\ge 1$. Finally, for all $\ell\in[1:\overline{L}-1]$, we have
\begin{align*}
\frac{\breve{R}_{\sf u\tdash dc}(R_{\sf c})}{R_{\sf u\tdash lb}(\Rcache)}  &=\frac{\frac{N-\omega_\ell}{1+\omega_\ell(1-1/N)}}{(1-(1-1/N)^\ell)(N-\ell\omega_\ell)} \\
&= \frac{\left[N+(\ell-N)\left(1-\frac{1}{N}\right)^\ell\right]\left[N+(\ell-N+1)\left(1-\frac{1}{N}\right)^\ell\right]}{\left[1-\left(1-\frac{1}{N}\right)^\ell\right]\left[N+\ell\left(1-\frac{1}{N}\right)^\ell\right]\left[N+(-N+1)\left(1-\frac{1}{N}\right)^\ell\right]} \\
&= \frac{\left[1-\left(1-\frac{1}{N}\right)^\ell+\frac{\ell}{N}\left(1-\frac{1}{N}\right)^\ell\right]\left[1-\left(1-\frac{1}{N}\right)^{\ell+1}+\frac{\ell}{N}\left(1-\frac{1}{N}\right)^\ell\right]}{\left[1-\left(1-\frac{1}{N}\right)^\ell\right]\left[1+\frac{\ell}{N}\left(1-\frac{1}{N}\right)^\ell\right]\left[1-\left(1-\frac{1}{N}\right)^{\ell+1}\right]} \\
&\le \frac{\left[1-\left(1-\frac{1}{N}\right)^\ell+\frac{\ell}{N}\left(1-\frac{1}{N}\right)^\ell\right]\left[1-\left(1-\frac{1}{N}\right)^{\ell+1}+\frac{\ell}{N}\left(1-\frac{1}{N}\right)^\ell\right]}{\left[1-\left(1-\frac{1}{N}\right)^\ell\right]\left[1-\left(1-\frac{1}{N}\right)^{\ell+1}\right]} \\
&= 1 + \frac{\frac{\ell}{N}\left(1-\frac{1}{N}\right)^\ell}{1-\left(1-\frac{1}{N}\right)^{\ell}}+\frac{\frac{\ell}{N}\left(1-\frac{1}{N}\right)^\ell}{1-\left(1-\frac{1}{N}\right)^{\ell+1}}+ \frac{\left(\frac{\ell}{N}\left(1-\frac{1}{N}\right)^\ell\right)^2}{\left[1-\left(1-\frac{1}{N}\right)^{\ell}\right]\left[1-\left(1-\frac{1}{N}\right)^{\ell+1}\right]} \\
&\le \left(1 + \frac{\frac{\ell}{N}\left(1-\frac{1}{N}\right)^\ell}{1-\left(1-\frac{1}{N}\right)^{\ell}}\right)^2 \\
&= \left(1 + \frac{\frac{\ell}{N}\alpha_N^{-\ell/N}}{1-\alpha_N^{-\ell/N}}\right)^2 \\
&\le \sup_{z\in(0,1]} \left(1 + \frac{z}{e^{z\ln \alpha_N}-1}\right)^2 \\
&\stackrel{(a)}{=} \left(1 + \frac{1}{\ln \alpha_N}\right)^2 \\
&= \left(1 - \frac{1}{N\ln(1-1/N)}\right)^2 \\
&\le 4,
\end{align*}
where $\alpha_N = (1-1/N)^{-N}$ and $(a)$ follows since $\frac{z}{e^{az}-1}$ is a decreasing function for all $a>0$.
This concludes the proof of Theorem~\ref{thm:gap-ergodic}.

\subsection{Proof of Theorem~\ref{thm:sb-gap}} \label{app:sb-gap}
Recall the definition of $\breve{R}_{\sf MN}(\Rcache)$, which is defined as the convexified bound in~\eqref{eq:MN-dec}.
If $N=1$, it can be easily checked that $\breve{R}_{\sf MN}(\Rcache)=1-\Rcache=R^\star_{\sf u\tdash ave}(\Rcache)$. For $N\in\{2,3,4\}$, we have
\begin{align*}
\frac{\breve{R}_{\sf MN}(\Rcache)}{R^\star_{\sf u\tdash ave}(\Rcache)} &\le \frac{N-\Rcache}{\max_{\ell\in[1:L]}(1-(1-1/N)^\ell)(N-\ell\Rcache)} \\
&\stackrel{\ell=1}{\le} \frac{N-\Rcache}{(1-(1-1/N))(N-\Rcache)} \\
&= N < 4.7.
\end{align*}
For the rest of analysis, we assume that $N\ge 5$.
To facilitate the gap analysis, we consider the following relaxed upper bound of~\eqref{eq:MN-dec}:
\begin{align*}
\breve{R}_{\sf MN}(\Rcache) &\le (N-\Rcache) \cdot  \min\left\{\frac{1}{\Rcache},1\right\} \\
&=: R_{\sf upper}(\Rcache),
\end{align*}
for all $\Rcache \in(0,N]$, and we define $R_{\sf upper}(0) := \min\{L,N\}$. We remark that $R_{\sf upper}(\Rcache)$ is quite suboptimal as an upper bound and is not continuous at $\Rcache=0$ when $L<N$. However, the corresponding convexified bound $\breve{R}_{\sf upper}(\Rcache)$ is sufficient for our analysis. On the other hand, we consider the following relaxed lower bound
\begin{align*}
R_{\sf u\tdash ave}^\star(\Rcache) &\ge \max_{\ell\in[\min\{L,\lceil N/4\rceil\}]}(1-(1-1/N)^\ell)(N-\ell\Rcache)^+ \\
&=: R_{\sf lower}(\Rcache).
\end{align*}
Since
\begin{align}
R_{\sf lower}(\Rcache) \le R_{\sf u\tdash ave}^\star(\Rcache) \le R_{\sf u\tdash wc}^\star(\Rcache) \le \breve{R}_{\sf MN}(\Rcache)\le \breve{R}_{\sf upper}(\Rcache), \label{eq:rate-relation}
\end{align}
it suffices to show
\begin{align*}
\frac{\breve{R}_{\sf upper}(\Rcache)}{R_{\sf lower}(\Rcache)} < 4.7, \quad \Rcache\in[0,N).
\end{align*}
For notational convenience, we denote $\overline{L} = \min\{L,\lceil N/4\rceil\}$ and $\kappa = \min\{L,N/4\}$.
Note that the lower bound $R_{\sf lower}(\Rcache)$ is an intersection of half planes. The corner points of $R_{\sf lower}(\Rcache)$ are characterized by the set $\Omega = \{(\omega_\ell,R_{\sf lower}(\omega_\ell)):\ell\in[0:\overline{L}]\}$, where
\begin{align*}
\omega_\ell &= \begin{cases}
N & \text{ if } \ell=0, \\
\frac{N\left(1-\frac{1}{N}\right)^\ell}{N+(\ell+1-N)\left(1-\frac{1}{N}\right)^\ell} & \text{ if } \ell\in[\overline{L}-1], \\
0 & \text{ if } \ell=\overline{L}.
\end{cases}
\end{align*}
We note that for all $\ell\in[\overline{L}-1]$, the two lines
\begin{IEEEeqnarray*}{rCl}
y &=& (1-(1-1/N)^\ell)(N-\ell x),  \\
y &=& (1-(1-1/N)^{\ell+1})(N-(\ell+1) x)
\end{IEEEeqnarray*}
intersect at $x=\omega_\ell$.

\begin{figure}[t!]
\begin{center}
\includegraphics[scale=0.5]{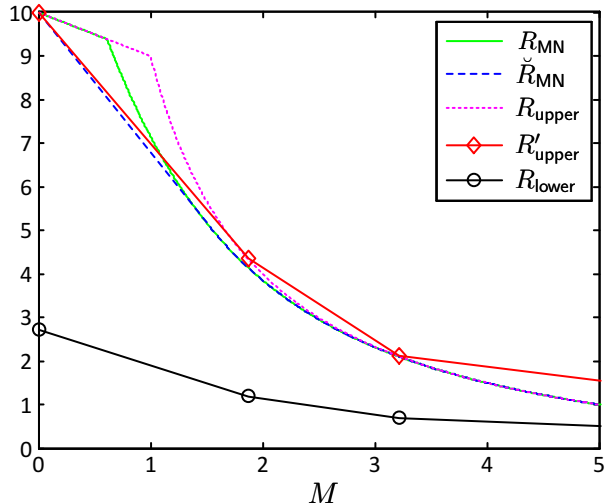}
\end{center}
\vspace{-0.2cm}
\caption{Plots of various bounds for $(K,N)=(15,10)$ and $\Rcache\in[0,5]$.}
\label{fig:N10K15}
\vspace{-0.2cm}
\end{figure}


Next, we relax the upper bound $\breve{R}_{\sf upper}(\Rcache)$ by the following piecewise-linear bound resulting from $\{\omega_\ell:k\in[0:\overline{K}]\}$:
\begin{IEEEeqnarray*}{ll}
& R'_{\sf upper}(\Rcache) \\
&:= (1-\theta)R_{\sf upper}(\omega_{\ell}) + \theta R_{\sf upper}(\omega_{\ell-1}),
\end{IEEEeqnarray*}
where $\Rcache = (1-\theta)\omega_{\ell} + \theta\omega_{\ell-1}$ for some $\theta\in [0,1)$, $\ell\in[1:\overline{L}]$. Note that $R'_{\sf upper}(\Rcache)
= R_{\sf upper}(\Rcache)$ for all $\Rcache\in\{\omega_\ell:\ell\in[0:\overline{L}]\}$. In Figure \ref{fig:N10K15}, we provide an example with $(K,N)=(15,10)$ summarizing the various bounds used in the analysis.

Then, for each segment $[\omega_{\ell},\omega_{\ell-1})$, $k\in[1:\overline{L}]$, the ratio $\frac{R'_{\sf upper}(\Rcache)}{R_{\sf lower}(\Rcache)}$ is a linear-fractional function with respect to $\Rcache$, and thus it is quasiconvex \cite{Boyd--Vandenberghe2004}. As  noted before, a quasiconvex function has the property that the value of the function on a segment does not exceed the maximum of its values at the endpoints. Thus, it suffices to check whether $\frac{R'_{\sf upper}(\Rcache)}{R_{\sf lower}(\Rcache)}< 4.7$ for all $\Rcache\in\{\omega_0,\omega_1,\cdots,\omega_{\overline{K}}\}$.

First, it is clear that we have
\begin{align*}
\lim_{\Rcache\to N^-}\frac{R'_{\sf upper}(\Rcache)}{R_{\sf lower}(\Rcache)} &= 1.
\end{align*}
Next, we have
\begin{align*}
\frac{R'_{\sf upper}(0)}{R_{\sf lower}(0)} &= \frac{\min\{L,N\}}{(1-(1-1/N)^{\overline{L}})N} \\
&\le \frac{4\kappa/N}{1-(1-1/N)^{\kappa}} \\
&\stackrel{(a)}{\le} 4\cdot \frac{\kappa/N}{1-e^{-\kappa/N}} \\
&\stackrel{(b)}{\le} \frac{1}{1-e^{-1/4}} \approx 4.521,
\end{align*}
where $(a)$ follows since $(1-1/z)^z \le e^{-1}$ for all $z > 1$ and $(b)$ follows since $\psi(z)=\frac{z}{1-e^{-z}}$ is an increasing function and $\kappa/N\le 1/4$.

As for $\ell\in[\overline{L}-1]$, we have
\begin{align*}
\frac{R'_{\sf upper}(\omega_\ell)}{R_{\sf lower}(\omega_\ell)}
&\le\frac{\frac{N-\omega_\ell}{\omega_\ell}}{(1-(1-1/N)^\ell)(N-\ell\omega_\ell)} \\
&= \frac{\left[N+(\ell-N)\left(1-\frac{1}{N}\right)^\ell\right]\left[N+(\ell-N+1)\left(1-\frac{1}{N}\right)^\ell\right]}{\left[N\left(1-\frac{1}{N}\right)^\ell\right]\left[1-\left(1-\frac{1}{N}\right)^\ell\right]\left[N+(-N+1)\left(1-\frac{1}{N}\right)^\ell\right]} \\
&= \frac{\left[1-\left(1-\frac{1}{N}\right)^\ell+\frac{\ell}{N}\left(1-\frac{1}{N}\right)^\ell\right]\left[1-\left(1-\frac{1}{N}\right)^{\ell+1}+\frac{\ell}{N}\left(1-\frac{1}{N}\right)^\ell\right]}{\left(1-\frac{1}{N}\right)^\ell \left[1-\left(1-\frac{1}{N}\right)^\ell\right]\left[1-\left(1-\frac{1}{N}\right)^{\ell+1}\right]} \\
&= \frac{1}{\left(1-\frac{1}{N}\right)^\ell}\left[1 + \frac{\frac{\ell}{N}\left(1-\frac{1}{N}\right)^\ell}{1-\left(1-\frac{1}{N}\right)^{\ell}}+\frac{\frac{\ell}{N}\left(1-\frac{1}{N}\right)^\ell}{1-\left(1-\frac{1}{N}\right)^{\ell+1}}+ \frac{\left(\frac{\ell}{N}\left(1-\frac{1}{N}\right)^\ell\right)^2}{\left[1-\left(1-\frac{1}{N}\right)^{\ell}\right]\left[1-\left(1-\frac{1}{N}\right)^{\ell+1}\right]}\right] \\
&\le \frac{1}{\left(1-\frac{1}{N}\right)^\ell}\left(1 + \frac{\frac{\ell}{N}\left(1-\frac{1}{N}\right)^\ell}{1-\left(1-\frac{1}{N}\right)^{\ell}}\right)^2 \\
&\stackrel{(a)}{=} e^{z}\left(1 + \frac{1}{\ln(1-1/N)^{-N}}\frac{z}{e^{z}-1}\right)^2 \\
&\stackrel{(b)}{\le} e^{z}\left(1 + \frac{z}{e^{z}-1}\right)^2  \\
&\stackrel{(c)}{\le} \left. e^z\left(1 + \frac{z}{e^z-1}\right)^2 \right|_{z=-\frac{N}{4}\ln(1-1/N)},
\end{align*}
where $(a)$ follows by a change of variable $z=-\ell\ln(1-1/N)$, $(b)$ follows since $z\ge 0$ and $(1-1/N)^{-N}\ge e$ for all $N > 1$, $(c)$ follows since $\phi(z) = e^{z}\left(1 + \frac{z}{e^{z}-1}\right)^2$ is an increasing function$^{\S}$\footnote{$^{\S}$ The function $\phi(z) = e^{z}\left(1 + \frac{z}{e^{z}-1}\right)^2$, $z\ge0$, is an increasing function since  its first derivative is nonnegative.} and $z\le -\frac{N}{4}\ln(1-1/N)$ (since $\ell \le \overline{L}-1 \le N/4$). Finally, since $-N\ln(1-1/N)$ is a decreasing function of $N$ and $N\ge 5$, we have
\begin{align*}
\frac{R'_{\sf upper}(\omega_\ell)}{R_{\sf lower}(\omega_\ell)} &\le \left. e^z\left(1 + \frac{z}{e^z-1}\right)^2 \right|_{z=-\frac{5}{4}\ln(1-1/5)} \\
&= \left. \nu^\nu\left(1 + \frac{\nu\ln\nu}{\nu^\nu-1}\right)^2 \right|_{\nu=\frac{5}{4}} \\
& \approx  4.607.
\end{align*}

%
%
\section{Proof of Theorem~\ref{thm:sbe-centralized}} \label{app:sbe-centralized}
We show the achievable rate pairs for $R_{\mathsf{c}}=0, \frac{1}{L},\frac{2}{L},\ldots, N$. Let $r_n\in[0:L]$, $n\in[1:N]$ such that $\sum_{n=1}^Nr_n=LR_{\mathsf{c}}$.
For the cache encoding step, we reuse the choice of $V_\ell$ in~\eqref{eq:def-vell} based on the definition of $W^{(n)}_{\Tc_n}$ in~\eqref{eq:def-w}. For the auxiliary random variables $U_\Sc$, $\Sc\subseteq[1:L]$, $\Sc\neq\emptyset$,  we choose
\begin{align}
U_\Sc=\left(W^{(n)}_{\Tc_n}\cdot A_\Sc(\Tc_n, n): n\in[1:N], {\Tc_n}\subseteq[1:L], |{\Tc_n}|=r_n\right) \label{eq:sbe-ues}
\end{align}
where
\begin{align}
\Ac_{\Sc}(\Tc_n, n) = \mathbb{1}\left\{\Tc_n\subset \Sc\right\}\mathbb{1}\left\{|\Sc|=r_n+1\right\}\left(\prod_{j\in\Sc\setminus\Tc_n}\mathbb{1}\left\{Y_j=n\right\}\right) \label{eq:sbe-Aes}.
\end{align}

Note that the above choice of auxiliary random variables satisfy~\eqref{eq:comp-inner-cond}. Since
\begin{align*}
I(U_\Sc; X|V_\ell, Y=y, Q)&=H(U_\Sc|V_{\ell}, Y=y, Q) \\
&= \sum_{n=1}^N \sum_{\Tc_n: |\Tc_n|=r_n} H(W^{(n)}_{\Tc_n}\cdot A_\Sc(\Tc_n, n)|V_\ell, Y=y, Q)\\
&= \sum_{n=1}^N \sum_{\Tc_n: |\Tc_n|=r_n,\ell\not\in\Tc_n} H(W^{(n)}_{\Tc_n}\cdot A_\Sc(\Tc_n, n)|Y=y, Q)\\
&= \sum_{n=1}^N \sum_{\Tc_n: |\Tc_n|=r_n,\ell\not\in\Tc_n} \frac{1}{\binom{L}{r_n}} H(W^{(n)}_{\Tc_n}\cdot A_\Sc(\Tc_n, n)|Y=y, Q, Q_n=\Tc_n)\\
&= \sum_{n=1}^N \mathbb{1}\{|\Sc|=r_n+1\}\sum_{\substack{\Tc_n: |\Tc_n|=r_n,\\ \ell\not\in\Tc_n, \Tc_n\subset\Sc}} \frac{1}{\binom{L}{r_n}} \mathbb{1}\{{y_{\Sc\setminus\Tc_n}=n}\}\\
&\stackrel{(b)}{=} \sum_{n=1}^N \mathbb{1}\{|\Sc|=r_n+1\}\sum_{\substack{\Tc_n: |\Tc_n|=r_n,\\ \ell\not\in\Tc_n, \Tc_n\subset\Sc}} \frac{1}{\binom{L}{r_n}} \mathbb{1}\{{y_{\ell}=n}\}\\
&= \sum_{n=1}^N\mathbb{1}\{|\Sc|=r_n+1\} \binom{|\Sc|-1}{r_n} \frac{1}{\binom{L}{r_n}} \mathbb{1}\{{y_{\ell}=n}\}\\
&= \sum_{n=1}^N\mathbb{1}\{|\Sc|=r_n+1\} \binom{r_n+1-1}{r_n} \frac{1}{\binom{L}{r_n}} \mathbb{1}\{{y_{\ell}=n}\}\\
&= \sum_{n=1}^N\mathbb{1}\{|\Sc|=r_n+1\} \frac{1}{\binom{L}{r_n}} \mathbb{1}\{{y_{\ell}=n}\}\\
&= \mathbb{1}\{|\Sc|=r_{y_\ell}+1\} \frac{1}{\binom{L}{|\Sc|-1}},
\end{align*}
where step $(a)$ follows since $W^{(n)}_{\Tc_n}\in V_\ell$ for $\ell\in\Tc_n$ and $(b)$ follows since for $|\Sc|=|\Tc_n|+1$, $\ell\not\in\Tc_n$, and $\ell\in\Sc$, we have $\Sc\setminus\Tc_n=\{\ell\}$.
Thus,
\begin{align}
R_{\mathsf{u}}(\Rcache,y)&> \sum_{\Sc:|\Sc|>0}\max_{\ell\in\Sc}H(U_\Sc|V_{\ell}, Y=y, Q) \nn\\
&=\sum_{\Sc:|\Sc|>0}\max_{\ell\in\Sc} \mathbb{1}\{|\Sc|=r_{y_\ell}+1\} \frac{1}{\binom{L}{|\Sc|-1}}\nn\\
&=\sum_{\Sc:|\Sc|>0}\left(1- \prod_{\ell\in\Sc}\mathbb{1}\{r_{y_\ell} \neq |\Sc| -1\}\right) \frac{1}{\binom{L}{|\Sc|-1}}. \label{eq:wc-centralized}
\end{align}
This concludes the proof of the first part of Theorem~\ref{thm:sbe-centralized}.

Next, for the average rate--cache tradeoff,
\begin{align*}
\E_Y\left[R_{\mathsf{u}}(\Rcache,Y)\right]&> \sum_{y}p_Y(y)\sum_{\Sc:|\Sc|>0}\left(1- \prod_{\ell\in\Sc}\mathbb{1}\{r_{y_\ell} \neq |\Sc| -1\}\right) \frac{1}{\binom{L}{|\Sc|-1}}\\
&\stackrel{(a)}{=} \sum_{\Sc:|\Sc|>0}\left(1- \prod_{\ell\in\Sc}\P\{r_{Y_\ell} \neq |\Sc| -1\}\right) \frac{1}{\binom{L}{|\Sc|-1}}\\
&= \sum_{\Sc:|\Sc|>0}\left(1- \prod_{\ell\in\Sc}\left(1-\P\{r_{Y_\ell} = |\Sc| -1\}\right)\right) \frac{1}{\binom{L}{|\Sc|-1}}\\
&= \sum_{\Sc:|\Sc|>0}\left(1- \prod_{\ell\in\Sc}\left(1-\sum_{n=1}^N\P\{r_{n} = |\Sc| -1, Y_\ell=n\}\right)\right) \frac{1}{\binom{L}{|\Sc|-1}}\\
&= \sum_{\Sc:|\Sc|>0}\left(1- \prod_{\ell\in\Sc}\left(1-\sum_{n=1}^N\mathbb{1}\{r_{n} = |\Sc| -1\}p_n\right)\right) \frac{1}{\binom{L}{|\Sc|-1}}\\
&= \sum_{j=1}^L\sum_{\Sc:|\Sc|=j}\left(1- \prod_{\ell\in\Sc}\left(1-\sum_{n=1}^N\mathbb{1}\{r_{n} = j -1\}p_n\right)\right) \frac{1}{\binom{L}{j-1}}\\
&= \sum_{j=1}^L\sum_{\Sc:|\Sc|=j}\left(1- \left(1-\sum_{n=1}^N\mathbb{1}\{r_{n} = j -1\}p_n\right)^j\right) \frac{1}{\binom{L}{j-1}}\\
&= \sum_{j=1}^L\binom{L}{j}\left(1- \left(1-\sum_{n=1}^N\mathbb{1}\{r_{n} = j -1\}p_n\right)^j\right) \frac{1}{\binom{L}{j-1}}\\
&= \sum_{j=1}^L\frac{L-j+1}{j}\left(1- \left(1-\sum_{n=1}^N\mathbb{1}\{r_{n} = j -1\}p_n\right)^j\right) \\
&= \sum_{j=0}^{L-1}\frac{L-j}{j+1}\left(1- \left(1-\sum_{n=1}^N\mathbb{1}\{r_{n} = j\}p_n\right)^{j+1}\right) \\
&= \sum_{j=0}^{L-1}\frac{L-j}{j+1}\left(1- \left(1-\alpha_j\right)^{j+1}\right),
\end{align*}
where $\alpha_j=\sum_{n=1}^N\mathbb{1}\{r_{n} = j\}p_n$, and step $(a)$ follows since $p_{Y}(y)=\prod_{j=1}^L p_{Y_1}(y_j)$.
This concludes the proof for the average rate--cache tradeoff.

Finally, to prove Remark~\ref{rmk:recover-cent}, we choose $r_n=r$, $n\in[1:N]$, such that $r=L \Rcache /N$. Then from~\eqref{eq:wc-centralized},
\begin{align*}
R_{\mathsf{u}}(\Rcache,y)&> \sum_{\Sc:|\Sc|>0}\left(1- \prod_{\ell\in\Sc}\mathbb{1}\{r_{y_\ell} \neq |\Sc| -1\}\right) \frac{1}{\binom{L}{|\Sc|-1}}\\
&= \sum_{\Sc:|\Sc|>0}\mathbb{1}\{r = |\Sc| -1\}\frac{1}{\binom{L}{|\Sc|-1}}\\
&= \sum_{\Sc:|\Sc|=r+1}\frac{1}{\binom{L}{r}}\\
&= \binom{L}{r+1}\frac{1}{\binom{L}{r}}\\
&= \frac{L-r}{1+r}\\
&= \frac{L-L\Rcache/N}{1+L\Rcache/N}.
\end{align*}

%
%
\section{Proof of Theorem~\ref{thm:sbe-decentralized}}\label{app:sbe-decentralized}
Consider any cache rate $R_{\mathsf{c}}\in[0,N]$ and let $r_n\in[0,1]$ such that $\sum_{n=1}^N r_n =R_{\mathsf{c}}$.
We choose the auxiliary random variables in the following manner.
For the cache encoding step, we reuse the choice of  $Q$ and $V_\ell$ in~\eqref{eq:vell-decentralized}.
Let $\Tc_n=\{\ell: Q^{(n)}_\ell=1\}$.
On the other hand, for $\mathcal{S}\subseteq [1:L]$, $\Sc\neq \emptyset$, we set
\begin{align*}
U_\mathcal{S} = \left(X^{(n)}A_{\mathcal{S},n}:n\in[1:N]\right),
\end{align*}
where
\begin{align*}
A_{\Sc,n}=\mathbb{1}\{\Tc_n\subset\Sc, |\Tc_n|=|\Sc|-1, Y_{\Sc\setminus\Tc_n}=n\}.
\end{align*}

Then, for $\mathcal{S}\subseteq[1:L]$ and $\ell\in\mathcal{S}$, we have
\begin{align*}
&H(X^{(n)}\cdot A_{\mathcal{S},n}|V_\ell,Y=y,Q) \\
&= H(X^{(n)}\cdot A_{\mathcal{S},n}|(X^{(n)}\cdot Q_\ell^{(n)}),Y=y,Q)\\
&= \P\{Q_\ell^{(n)}=0\} H(X^{(n)}\cdot A_{\mathcal{S},n}|(X^{(n)}\cdot Q_\ell^{(n)}),Y=y,Q,Q_\ell^{(n)}=0) \\
&= \P\{Q_\ell^{(n)}=0\} \P\{\Tc_n\subset\Sc, |\Tc_n|=|\Sc|-1, y_{\Sc\setminus\Tc_n}=n|Q_\ell^{(n)}=0\} \\
&\stackrel{(a)}{=} \P\{Q_\ell^{(n)}=0\} \P\{\Tc_n\subset\Sc, |\Tc_n|=|\Sc|-1, y_\ell=n|Q_\ell^{(n)}=0\} \\
&= \P\{Q_\ell^{(n)}=0\} \mathbb{1}\{y_\ell=n\} \P\{\Tc_n\subset\Sc, |\Tc_n|=|\Sc|-1|Q_\ell^{(n)}=0\} \\
&= \P\{Q_\ell^{(n)}=0\} \mathbb{1}\{y_\ell=n\} \prod_{j\in S\setminus \{\ell\}}\mathbb{1}\{Q_j^{(n)} = 1\}\prod_{j\in S^c}\mathbb{1}\{Q_j^{(n)}=0\} \\
&= \P\{Q_\ell^{(n)}=0\} \mathbb{1}\{y_\ell=n\} \prod_{j\in\mathcal{S}\backslash\{\ell\}}
r_n \prod_{j\in\mathcal{S}^c} (1-r_{n}) \\
&= \mathbb{1}\{y_\ell=n\}\left(1-r_n\right)r_n^{|\mathcal{S}|-1}(1-r_n)^{|\mathcal{S}^c|},
\end{align*}
where $(a)$ follows since for $|\Sc|=|\Tc_n|+1$, $\Tc_n\subset \Sc$, $Q_\ell^{(n)}=0$, and $\Tc_n=\{\ell': Q^{(n)}_{\ell'}=1\}$, the condition $y_{\Sc\setminus\Tc_n}=n$ is equivalent to $y_\ell=n$.
Thus, it holds that
\begin{align*}
R_{\sf u}(\Rcache,y) &> \sum_{\mathcal{S}\subseteq[1:L]} \max_{\ell\in\mathcal{S}} H(U_{\mathcal{S}}|V_\ell,Y,Q) \\
&= \sum_{j=1}^L \sum_{\mathcal{S}:|\mathcal{S}|=j} \max_{\ell\in\mathcal{S}}\sum_{n=1}^N \mathbb{1}\{y_\ell=n\} \left(1-r_n\right)r_n^{|\mathcal{S}|-1}(1-r_n)^{|\mathcal{S}^c|} \\
&= \sum_{j=1}^L \sum_{\mathcal{S}:|\mathcal{S}|=j} \max_{\ell\in\mathcal{S}}\sum_{n=1}^N \mathbb{1}\{y_\ell=n\} r_n^{j-1}(1-r_n)^{L-j+1} \\
&= \sum_{j=1}^L \sum_{\mathcal{S}:|\mathcal{S}|=j} \max_{\ell\in\mathcal{S}} r_{y_\ell}^{j-1}(1-r_{y_\ell})^{L-j+1}.
\end{align*}
This concludes the proof of the first part of Theorem~\ref{thm:sbe-decentralized}.

Finally, to prove Remark~\ref{rmk:recover-decent}, we choose $r_n=r=\Rcache/N$. Thus,
\begin{align*}
R_{\sf u}(\Rcache,y) &>\sum_{j=1}^L \sum_{\mathcal{S}:|\mathcal{S}|=j} \max_{\ell\in\mathcal{S}} r_{y_\ell}^{j-1}(1-r_{y_\ell})^{L-j+1}\\
&= \sum_{j=1}^L \sum_{\mathcal{S}:|\mathcal{S}|=j}  r^{j-1}(1-r)^{L-j+1} \\
&= \frac{\left(1-r\right)}{r} \sum_{j=1}^L \binom{L}{j} r^{j}(1-r)^{L-j} \\
\label{eq:geo}
&= \frac{\left(1-r\right)}{r} (1-(1-r)^L)\\
&= \frac{\left(N-\Rcache\right)}{\Rcache} (1-(1-\Rcache/N)^L).
\end{align*}

%
%
\section{Proof of Corollary~\ref{cor:optimal-regime}} \label{app:optimal-regime}
For $R_{\sf c}\ge 0$, we first relax the lower bound~\eqref{eq:WLGouter} by fixing $\ell=1$ and get
\begin{align}
R^\star_{\mathsf{u}} (\Rcache) &\ge \sum_{n=1}^N (s_n(1)-s_{n+1}(1))\left(n- \Rcache\right)^+  \\
&\ge  p_N\left(N- \Rcache\right). \label{eq: upper-ell-1}
\end{align}

For the uniform request case, by choosing $r_1=\cdots=r_{N}=L-1$ for $\Rcache=N-N/L$ in Theorem~\ref{thm:inner-centralized}, we have
\begin{align*}
R^\star_{\mathsf{u}} (R_{\mathsf{c}})&\le \frac{1}{L}.
\end{align*}
By memory-sharing between $R^\star_{\mathsf{u}}(N)=0$, we have that for $\Rcache\in[N-N/L, N]$,
\begin{align*}
R^\star_{\mathsf{u}} (\Rcache) &\le p_N\left(N- \Rcache\right).
\end{align*}
Furthermore, for arbitrary requests, by choosing $r_1=\cdots=r_{N-1}=L$ and $r_N=L-1$ for $\Rcache=N-1/L$ in Theorem~\ref{thm:inner-centralized}, we have
\begin{align*}
R^\star_{\mathsf{u}} (N-1/L) &\le \frac{1}{1+r_N} p_N\\
&= \frac{p_N}{L}.
\end{align*}
By memory-sharing between $R^\star_{\mathsf{u}}(N)=0$, we have that for $\Rcache\in[N-1/L, N]$,
\begin{align*}
R^\star_{\mathsf{u}} (\Rcache) &\le p_N\left(N- \Rcache\right).
\end{align*}

\section{Proof of Proposition~\ref{prop:greedy}}\label{app:greedy}
We prove the proposition by induction. First, for $R_{\sf c}=0$, Algorithm \ref{alg:greedy} is initialized by the optimal value $R^\star_{\sf u}(0)$. Next, we assume that Algorithm \ref{alg:greedy} finds the minimum value of $R_{\sf u}(\Rcache)$ (the right hand side of \eqref{eq:inner-centralized}) when $R_{\sf c}=s/L$ for some $s\in[1:NL]$. Denote by $\rv_s^\star=(r_1^\star,\ldots, r_N^\star)$ the corresponding assignment in Algorithm \ref{alg:greedy} for $R_{\sf c}=s/L$. For $r\in[0:L]$ and $p\in[0,1]$, denote
\begin{align*}
\kappa(r,p) &= \sum_{j=1}^{L-r} \frac{j}{j+r} \binom{L-r}{j} p^j(1-p)^{L-r-j}.
\end{align*}
We observe that for $r\in[1:L]$ and $p\in[0,1]$
\begin{align*}
\kappa(r,p) &= \E_{Z}\left[\frac{Z}{Z+r}\right],
\end{align*}
where $Z\sim \text{Binom}(L-r,p)$, and
\begin{align*}
R(\mathbf{r}) = \sum_{n=1}^N \kappa(r_n, p_n).
\end{align*}
Since it will be clear from the context, we simply denote $\kappa(r_n)=\kappa(r_n, p_n)$.
Note that the induction hypothesis implies $R_{\sf u}(s/L)=R(\mathbf{r}^\star_s) $.
Then, for the case $R_{\sf c}=(s+1)/L$, assume an arbitrary $\rv=(r_1,\ldots, r_N)\in[0:L]^N$ such that $\sum_{n=1}^Nr_n=s+1$. Note that from the pigeonhole principle, there exists a component $j$ such that $r_j \ge r^\star_j+1$.
Let $\mathbf{1}_j$ be an all zero vector with the $j$th component replaced by $1$. Then, 
\begin{align*}
R(\mathbf{r}) &= R(\mathbf{r}-\mathbf{1}_j) - (\kappa(r_j-1) - \kappa(r_j)) \\
&\stackrel{(a)}{\ge} R(\rv^\star_s) - (\kappa(r_j-1) - \kappa(r_j)) \\
&\stackrel{(b)}{\ge} R(\rv^\star_s) - (\kappa(r_j^\star) - \kappa(r_j^\star+1)) \\
&\ge R(\rv^\star_s) - \max_{n\in[1:N]} (\kappa(r^\star_n) - \kappa(r^\star_n+1))\\
&\stackrel{(c)}{=} R(\rv^\star_{s+1}),
\end{align*}
where for convenience we define $\kappa(L+1)=0$, step $(a)$ follows from the fact that the element-wise sum of $\rv - \mathbf{1}_j$ is $s$ and from the induction hypothesis, 
step $(b)$ follows since for $r\in[1:L-1]$,
\begin{align} \label{eq:cvx}
\kappa(r-1)-\kappa(r) \ge \kappa(r)-\kappa(r+1),
\end{align}
and that $r_j \ge r^\star_j+1$, and step $(c)$ follows from the incremental assignment of $\rv_{s+1}^\star$ from $\rv_s^\star$ in Algorithm \ref{alg:greedy}.
It remains to prove~\eqref{eq:cvx} which we show in the following.
First, we consider the case $r=1$. Let $Z\sim \text{Binom}(L-2,p)$ and $A\sim \text{Bern}(p)$. Assume that $Z$ and $A$ are independent. Then, we have $Z+A\sim \text{Binom}(L-1,p)$ and thus
\begin{align*}
2\kappa(1) &= 2\E\left[\frac{Z+A}{Z+A+1}\right] \\
&= 2\E\left[\E\left[\left.\frac{Z+A}{Z+A+1}\right|A\right]\right] \\
&= 2p\E\left[\frac{Z+1}{Z+2}\right] + 2(1-p)\E\left[\frac{Z}{Z+1}\right] \\
&= \kappa(2) + \E\left[\frac{(2p-1)Z+2p}{Z+2}\right] + \E\left[\frac{2(1-p)Z}{Z+1}\right] \\
&= \kappa(2) + 1 - 2(1-p) \E\left[\frac{1}{(Z+1)(Z+2)}\right] \\
&= \kappa(2) + 1 - 2(1-p) \left[\frac{1}{2p^2\binom{L}{2}}\sum_{j=2}^L\binom{L}{j}p^j(1-p)^{L-j}\right] \\
&\le \kappa(2) + 1 - 2(1-p) \left[\frac{1}{2p^2\binom{L}{2}}\binom{L}{2}p^2(1-p)^{L-2}\right] \\
&= \kappa(2) + 1 - (1-p)^{L-1} \\
&\le \kappa(2) + \kappa(0).
\end{align*}
Next, we consider the case $r\ge 2$. Let $U\sim \text{Binom}(L-r-1,p)$ and $A,B\sim \text{Bern}(p)$. Assume that $U,A,B$ are independent. Denote $V=U+A$. Then, we have
$V\sim \text{Binom}(L-r,p)$, $V+B\sim \text{Binom}(L-r+1,p)$, and
\begin{align*}
& \kappa(r-1) - \kappa(r) \\
&= \E\left[\frac{V+B}{V+B+r-1}\right] - \E\left[\frac{V}{V+r}\right] \\
&= p\E\left[\frac{V+1}{V+r}\right] + (1-p)\E\left[\frac{V}{V+r-1}\right] - \E\left[\frac{V}{V+r}\right] \\
&= p\E\left[\frac{1}{V+r}\right] + (1-p)\E\left[\frac{V}{V+r-1}\right] -(1-p) \E\left[\frac{V}{V+r}\right] \\
&= p\E\left[\frac{1}{V+r}\right] + (1-p)\E\left[\frac{V}{(V+r-1)(V+r)}\right] \\
&\ge p\E\left[\frac{1}{U+r+1}\right] + (1-p)\E\left[\frac{U}{(U+t)(U+r+1)}\right] \\
&= p\E\left[\frac{1}{U+r+1}\right] + (1-p)\E\left[\frac{U}{U+r}\right]-(1-p)\E\left[\frac{U}{U+r+1}\right] \\
&= p\E\left[\frac{U+1}{U+r+1}\right] + (1-p)\E\left[\frac{U}{U+r}\right]-\E\left[\frac{U}{U+r+1}\right] \\
&= \E\left[\frac{U+A}{U+A+r}\right] - \E\left[\frac{U}{U+r+1}\right] \\
&= \kappa(r) - \kappa(r+1).
\end{align*}

\bibliographystyle{IEEEtran}
\newcommand{\noopsort}[1]{}

\end{document}